\documentstyle[sidef]{l-aa-ps}
\input{psfig}
  \def\hk{$(H-K)$}
\def\jh{$(J-H)$}  

\begin{document}
\thesaurus{08 (08.02.3; 08.02.6; 08.06.2; 08.16.5; 10.15.2 Sco OB2; 10.15.2 Chamaeleon)}
\title{Multiplicity among T\,Tauri stars in OB and T associations\thanks{Based
on observations obtained at the European Southern Observatory, La Silla}}
\subtitle{Implications for binary star formation}
\author{Wolfgang Brandner \inst{1}, Juan M. Alcal\'a \inst{2}, Michael Kunkel \inst{1}, Andrea Moneti \inst{3}, and Hans Zinnecker \inst{1}}
\institute{
Astronomisches Institut der Universit\"at W\"urzburg, Am Hubland, 
97074 W\"urzburg, Germany\\ brandner@astro.uni-wuerzburg.de, kunkel@astro.uni-wuerzburg.de, hans@astro.uni-wuerzburg.de
\and
Max Planck Institut f\"ur extraterrestrische Physik, 85740 Garching, Germany\\
jmae@mpe-garching.mpg.de
\and
ESTEC -- SAI, PO Box 299, 2200 AG Noordwijk, The Netherlands\\ amoneti@iso.estec.esa.nl}
\offprints{Wolfgang Brandner}
\date{Received 21 April 1995, accepted date}
\maketitle
%\maintitlerunninghead{Multiplicity among T\,Tauri stars in OB and T associations}
%\authorrunninghead{Brandner et al.}
\markboth{Multiplicity among T\,Tauri stars: Implications on binary formation}{Brandner et al.}
\begin{abstract} 
We present first results of a survey for companions among X-ray
selected pre-main sequence stars, most of them being
weak-line T Tauri stars (WTTS). These T Tauri stars have been identified
in the course of optical follow-up observations of sources from the
ROSAT All Sky Survey associated with star forming regions. 
The areas surveyed include the T associations of Chamaeleon and Lupus
as well as Upper Scorpius, the latter
being part of the Scorpius Centaurus OB association (Sco OB 2).

Using SUSI at the NTT under subarcsec seeing conditions we observed 
195 T Tauri stars through a 1$\mu$m (``Z'') filter and identified companions to 
31 of them (among these 12 subarcsec binaries).
Based on statistical arguments we conclude that almost all of them are 
indeed physical (i.e. gravitationally bound) binary or multiple systems. 
For 10 systems located in Upper Scorpius and Lupus, we additionally obtained 
spatially resolved near-infrared photometry in the J, H, and K bands
with the MPIA 2.2m telescope at ESO, La Silla. 
The near-infrared colours of the secondaries are consistent with those of 
dwarfs and are
clearly distinct from those of late type giant stars. 
Based on astrometric measurements of some binaries we show that the
components of these binaries are common proper motion pairs, very likely
in a gravitationally bound orbit around each other.

We find that the overall binary frequency among T Tauri stars in a range of
separations between 120 and 1800 AU is in
agreement with the binary frequency observed among main sequence stars
in the solar neighbourhood. However, we note that within individual regions
the spatial distribution of binaries -- within a distinct range of separation --
 is non-uniform. In particular, 
in Upper Scorpius, WTTS in the vicinity of early type stars seem to be almost
devoid of multiple systems, whereas in another area in Upper Scorpius half 
of all
WTTS have a companion in a range of separation between 0\farcs7 and 3\farcs0.
Furthermore, we find no preponderance of systems with {\it large} brightness 
differences between primary and companion stars (median $\Delta$Z = 
1\fm0 \dots 1\fm5).

We conclude that binarity is established very early in stellar evolution,
that the orbital parameters of {\it wide} binaries (a $\ge$ 120AU) remain
virtually unchanged during their pre-main sequence evolution, and that these 
{\it wide} binaries were formed either through collisional fragmentation
or fragmentation of rotating filaments.


\end{abstract}

\section{Introduction}

Prior studies of X-ray sources associated with star forming regions
based on data obtained by the EINSTEIN satellite
led to the discovery of a new class of low-mass pre-main sequence
stars, the so called weak-line T Tauri stars (WTTS, Feigelson
\& Kriss 1981, Walter \& Kuhi 1981, Montmerle
et al.\ 1983).
However, the sky coverage of the EINSTEIN pointings is rather inhomogeneous
and thus statistically sound studies of large samples of T Tauri stars are
severely hampered by selection effects. 
The ROSAT All Sky Survey (``RASS'') with its uniform sky coverage
provides a much more homogeneous sample and constitutes a unique database
of X-ray sources in all nearby star forming regions.
The number of known T Tauri stars (TTS)
has grown considerably in the course of optical follow-up observations of
RASS sources associated with nearby star forming regions. 
Alcal\'a et al.\ (1995), Wichmann (1994), and Kunkel (1995) identified the 
optical
counterparts to about 500 RASS sources as T Tauri stars, most of them WTTS.

%Cha 81, Ori 18, Tau 76, Lup 113, Sco 112(6)?

We now have a large and relatively unbiased
sample of low-mass pre-main-sequence stars which is open for statistical
studies. In our present study we aim at detecting companions to these
T Tauri stars.

Most solar-type main sequence stars are members of binary or 
multiple systems (e.g. Duquennoy \& Mayor 1991) 
and therefore studying star formation means to a large 
extent studying the formation of binary systems.
Understanding binary formation holds the key for a more general understanding
of star formation.

Yet, it is still unknown to what an extent the binary frequency is the same
in different kinds of star forming regions like T associations, 
OB associations, or dense stellar aggregates like the Trapezium Cluster
in Orion. There may also be systematic differences between WTTS and
classical T Tauri stars (CTTS).

Recent studies among CTTS and WTTS in the Taurus-Auriga T association 
(Simon et al.\ 1992, Simon 1992, Ghez et al.\ 1993, Leinert et al.\ 1993, 
Richichi et al.\ 1994)
reveal that as a lower limit 51\% of all T Tauri stars 
in that particular region have at least one companion in a range of
separations between 1.8 AU to 1800 AU (for a comprehensive
compilation of all known binary and multiple systems in Taurus,
see Mathieu, 1994). Thus the
binary frequency in Taurus-Auriga is enhanced in comparison to main sequence
stars in the solar neighbourhood, where only 40\% of all G type stars
show companions in the same range of separations (Duquennoy \& Mayor, 1991,
see also Fig.1 in Mathieu, 1994). The implications of this result are still
unclear, and it is important to see if the excess holds true for other
regions.

Ghez et al.\ (1993) find ``(...) that the WTTS binary star distribution is
enhanced at the closer separations ($<$ 50AU) relative to the CTTS binary
star distribution'', a result which is disputed by Leinert et al.\ (1993).  
Leinert et al.\ find no significant difference in the binary
frequency and the distribution of separations among CTTS and WTTS.
This shows that even in Taurus--Auriga, the best studied region, the results 
are still unclear. Results for other regions (Ghez et al.\ 1993, 
Reipurth \& Zinnecker 1993) are not statistically
sound because of (i) smaller sample sizes, and (ii) more limited range
of separations. Clearly, one needs a larger sample size in order to
establish unambiguous results.

Reipurth \& Zinnecker (1993) surveyed about
230 CTTS in the southern T associations of Chamaeleon, Lupus, and Ophiuchus
and identified $16\% \pm 3\%$ CTTS to be binary or triple systems 
in a range of separations between 150AU and 1800AU. Mathieu (1994) has
pointed out that this value is in agreement with the binary frequency
among main sequence stars in the same range of separations 
within the statistical uncertainties ($12\% \pm 3\%$).

Our present study is a continuation of the work by 
Reipurth \& Zinnecker. We define two goals: first, we want to increase
the sample size by extending our survey to all the newly detected T Tauri 
stars in the above mentioned T associations in order to improve our
statistics. 
In doing so, we want to look for systematic differences in binary 
frequency among CTTS and WTTS, since WTTS might outnumber CTTS by a
factor of 2 to 10 (Walter et al.\ 1994, Sterzik et al.\ 1995), and hence
dominate the statistics.
Second, by including also T Tauri stars in the 
Scorpius--Centaurus OB association, we can for the first
time compare the low-mass binary content
of T  and OB associations.

\section{The regions under study and selection of programme stars}

Throughout the whole paper we are using the two terms 
``WTTS'' and ``X-ray selected TTS'' as synonyms. The same holds true for 
``CTTS'' and ``H$\alpha$ selected TTS''. There are of course CTTS which
are visible in X-rays and vice versa, but only $\approx$ 5\%
of all TTS fall into this category. 
We note also that the distinction
between CTTS and WTTS is by no means sharp. H$\alpha$ equivalent widths
are varying and sometimes E$_{H\alpha}$ = 0.5nm or E$_{H\alpha}$ = 1.0nm
is used to distinguish between both classes. 

In addition to the ROSAT All Sky Survey also pointed observations of star 
forming regions 
were carried out by ROSAT. In contrast to the relatively short exposure
times in the RASS (300--1500s), 
the pointed observations go much deeper
(up to 35000s), and thus are more complete with respect to the low-mass PMS 
population of the star forming regions. However, they lack the spatial
homogeneity of the RASS. 
Therefore, in the first part of our survey,
we concentrated on the RASS sources, but -- especially in 
Chamaeleon
-- also TTS detected in pointed observations were observed.

The Scorpius--Centaurus OB association (Sco OB 2) at a distance of $\approx$ 
150 pc is the most nearby OB association (Blaauw 1964). It consists of the 
four subgroups
Lower 
Centaurus - Crux, Upper Centaurus - Lupus, Upper Scorpius, 
and Ophiuchus.
Their ages are 11 Myrs, 13 Myrs, 7 Myrs, and 2-6 Myrs, respectively
(Blaauw 1991). The only O-type star still left in the entire region is the 
runaway star $\zeta$ Oph (O9.5). 

Blaauw (1991) derived a kinematical age
of 10$^6$yrs for $\zeta$ Oph. He concluded that the last SN
explosion in Upper Scorpius, which gave ``birth'' to $\zeta$ Oph as a 
runaway star, happened about 10$^6$yrs ago.
Supernova explosions are thought to have initiated 
a sequential star formation process, spreading out from Upper Centaurus over 
Upper Scorpius to
Ophiuchus, where deeply embedded young stars can still be found in the
center of molecular cloud cores (e.g., Wilking et al.\ 1989).

In Upper Scorpius, the last SNe very likely initiated
the star formation process in the dark clouds around $\rho$ Ophiuchi.
Kunkel (1995) identified about 100
T Tauri stars in the course of optical follow-up studies of 
sources from the RASS in the Upper Scorpius region itself.
In a parallel programme we carried out an 
H$\alpha$ objective
prism survey using the ESO Schmidt-telescope in order to detect and identify
CTTS. In this paper, however, we will only report on
the X-ray selected T\,Tauri stars since the identification of the
H$\alpha$ selected TTS has not yet been finished. We will report on these
stars in a subsequent paper.

Situated in the immediate neighbourhood of Sco OB2 is the Lupus T association.
CTTS are known in association with the four dark clouds Lupus 1 to 4
(Schwartz 1977). Wichmann (1994) has identified about 115 new TTS as
counterparts to ROSAT sources.
In the Scorpius and Lupus region we surveyed 74 of the 112 TTS 
identified by Kunkel (1995) on the basis of the RASS. 
Six of these stars have also
been identified by Wichmann (1994) as new TTS. None of the remaining
107 new TTS from the list of Wichmann has been surveyed so far.

We additionally observed all the CTTS candidates in Lupus 4 and in the 
Norma region from the list of Schwartz (1977). Because of uncertainties
in the association membership, their status as TTS at all (see the
comments by Schwartz 1977, Table 8 \& 9), and distance estimates
to the Norma cloud ranging from 200pc to 940pc 
(see the discussion by Reipurth \& Graham 1991)
 we did not include these stars into our
final statistics (Table \ref{mult}). The only binary found among these
objects is Sz 128 (cf. Table \ref{cross}).

The Chamaeleon T association consists of the three dark clouds
Cha I, II, and III (following the designations of Schwartz 1977)
at a distance between 140pc and 170pc. CTTS are only 
known close to Cha I and Cha II (Schwartz 1977, Whittet et al.\ 1987, 
Hartigan 1993). 
Follow-up studies of ROSAT observations led to
the identification of about 100 new WTTS (Alcal\'a et al.\ 1995, Huenemoerder 
et al.\ 1994) spread over an area
much wider than that covered by the molecular clouds themselves.
All of the 95 ROSAT sources in Chamaeleon, to which TTS have
been identified as optical counterparts by Alcal\'a et al.\ (1995)
 and/or Huenemoerder et al.\ (1994), have been searched for companions. 
In the cases where two TTS fell within the ROSAT
error circle, only the one closest to the X-ray 
position was observed (RXJ1108.8-7519, RXJ1109.9-7629, RXJ1301.0-7654).
In addition, we observed the 26 H$\alpha$ selected TTS from the list
of Hartigan (1994). Together with our previous observations
(Brandner 1992, Reipurth \& Zinnecker 1993) we have surveyed all 195 up to
now identified TTS in Chamaeleon.

All the above mentioned optical follow-up studies of ROSAT sources have in 
common that they are magnitude limited with the
faintest stars having m$_V \approx$ 15$^{\rm m}$ -- 16$^{\rm m}$. Deeper optical
surveys suffer from the growing contamination by extragalactic objects.
But also the ROSAT All Sky Survey itself is magnitude limited. There are WTTS too faint
in X-rays to be discovered by ROSAT.
Both indicates an incompleteness of our current
survey towards the low-mass end of the IMF in these regions.

Finding charts and additional information (e.g., L$_{\rm bol}$ and spectral
types) for the TTS in Chamaeleon can be found in Alcal\'a (1994),
 Alcal\'a et al.\ (1995), Feigelson 
et al.\ (1993), Hartigan (1993), and Huenemoerder et al.\ (1994),
for the TTS in Upper Scorpius in Kunkel (1995), and for the TTS in
Lupus in Wichmann (1994).

\section{Observations and data reduction} 
\subsection{1$\mu$m imaging (NTT)}

\begin{table}
\caption{\label{obs}Journal of observations}  
\begin{center}
\begin{tabular}{cccc}
telescope/instr.& date & filter & seeing\\ \hline
NTT/SUSI     &20.3.1994 & Z (1$\mu$m) & 0\farcs45--0\farcs80 \\   
NTT/SUSI     & 17.5.1994 & Z (1$\mu$m)& 0\farcs60--1\farcs10 \\   
MPIA 2.2m/IRAC2B  &17.6.1994 & JHK & 0\farcs8--1\farcs3 \\     
NTT/SUSI     & 6./7.2.1995 & Z (1$\mu$m)& 0\farcs60--1\farcs00 \\   
\hline
\end{tabular}
\end{center}
\end{table}

\begin{figure}
\centerline{\psfig{figure=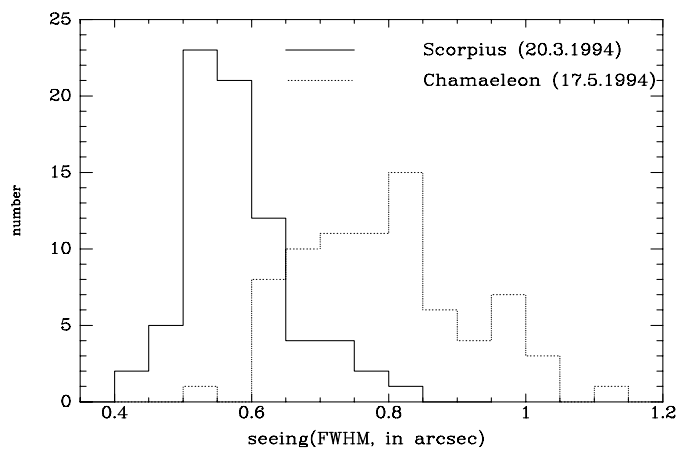,width=9.5cm}}
\caption{\label{see_sco}Histogram of the distribution of seeing values
(FHWM) for the observations of X-ray selected TTS in Scorpius/Lupus
(solid line)
and in Chamaeleon (dotted line) with SUSI/NTT. The typical
exposure time was 30s, except for the two exposures with a seeing value of
0\farcs44, when the exposure time was 10s.}
\end{figure}

The search for close companions was carried out in four nights at the NTT 
under sub-arcsec seeing conditions. 
In 1994 March 20 we observed X-ray selected
TTS in the Upper Scorpius region and in Lupus from the list of Kunkel (1995), 
and
in 1994 May 17 we observed X-ray selected TTS in the Chamaeleon region from the
list of Alcal\'a et al.\ (1995). Finally, in February 1995 we observed again 
the Chamaeleon
region, but this time X-ray selected TTS from the list of Huenemoerder et al.\
(1994), and H$\alpha$ selected TTS from the list of Hartigan (1993)
as well as CTTS in Lupus 4 and Norma (Schwartz 1977).
The observations were obtained with SUSI, the SUperb Seeing Imager, which
is equipped with a 1K TEK CCD (ESO \#25) and has an image scale of
0\farcs13/pixel. An intermediate band 1$\mu$m (``Z'') filter (ESO \#759 or \#760,
FWHM 50nm)
was used. Observations in the far-red benefit from the better seeing 
compared to the visual and facilitate the detection
of companions either embedded or of very late spectral type.
The NTT is equipped with an active optics system which allows us to
check (``image analysis'') and, if necessary, correct the image quality 
online (by adjusting 
the shape of the primary mirror and the position and orientation of the 
secondary mirror).

\begin{table*}
\caption{\label{cross}Detected pre-main sequence binaries and their measured parameters}
\begin{center}
\begin{minipage}{180mm}
\begin{tabular}{llllrrrccl}
\hline
ROSAT design. &$\alpha$(2000) & $\delta$(2000) & alias & separation&PA& $\Delta$Z&  SpT & m$_V$ & E$_{H\alpha}$\\
\hline
RXJ1039.5-7538\footnote{This star has already been observed in 1992 with SUSI/NTT. The values reported here are from Brandner (1992). }
&10 39 31.8&-75 37 56  & HD 92727 &5\farcs04$\pm$0\farcs05 &351\fdg2$\pm$1\fdg0&$\approx$ 1.5 &    G0&
9.3+10.1 & W \\
RXJ1101.3-7627\footnote{Huenemoerder et al.\ (1994) already noted the
 binary nature of this star}&11 01 18.2&-76 26 59&CHXR 9C   &   0\farcs96$\pm$0\farcs01 &268\fdg9$\pm$1\fdg0&0.04 &     &12.2(R) &W\\
&11 02 34.1&-77 29 12&Hn 1&   0\farcs68$\pm$0\farcs05 &77\fdg2$\pm$2\fdg0&1.52 &     &14.6(R) &\\
RXJ1108.1-7742\footnote{Brandner (1992) identified the companion based on direct 
imaging (SUSI/NTT). The parameters reported there are: sep.\ 0\farcs69,
PA $179^\circ$, and $\Delta$Z=0.38.}
&11 08 02.7&-77 42 28  &    VW Cha, Sz 24 & 0\farcs72$\pm$0\farcs03 &178\fdg6$\pm$1\fdg0&0.25 &    K5&
12.6 & C \\
RXJ1108.2-7728&11 07 55.7&-77 27 24&CHX 10    &   1\farcs87$\pm$0\farcs05 &119\fdg8$\pm$1\fdg0&1.06 & K6  &13.60 &W\\
&11 09 19.3&-76 30 27&Hn 9\footnote{Hartigan (1993) reports Hn 8 to be a binary
with sep.\ $<1''$ and a PA $\approx$ 210$^\circ$. This we could not confirm, 
but we found Hn 9 to be a binary with similar properties.} &   0\farcs92$\pm$0\farcs05 &216\fdg1$\pm$1\fdg0&1.64 &     &17.0(R) &\\
&11 14 25.9&-77 33 06&Hn 21  &   5\farcs54$\pm$0\farcs05 &71\fdg8$\pm$1\fdg0&0.66 &     &16.5+17.1(R) &\\
&11 18 20.2&-76 21 59&CHXR 68$^b$   &   4\farcs46$\pm$0\farcs05 &215\fdg3$\pm$1\fdg0&1.50 &     &12.2(R) &W\\
RXJ1150.9-7411&11 50 44.9&-74 11 13&          &   0\farcs91$\pm$0\farcs03 &106\fdg7$\pm$1\fdg0&1.38 &    M4 &14.43 &W\\
RXJ1207.9-7555\footnote{We reobserved this close binary in Feb. 1995 and found the companion this time to be marginally brighter than the primary ($\Delta$Z = -0.07).}&12 07 53.5&-75 55 14&   CPD-75 783&   0\farcs64$\pm$0\farcs03 &8\fdg2$\pm$2\fdg0&0.01&K2&  
9.9 &W\\
RXJ1243.1-7458&12 42 52.0&-74 58 46&           &   2\farcs58$\pm$0\farcs03 &261\fdg6$\pm$1\fdg0&1.78& M3&15.10 &W\\
RXJ1301.0-7654\footnote{This star was found to be a spectroscopic 
binary by Covino et al.\ (1995, in preparation). Thus, if this 
visual binary is a physical system, this star forms a hierarchical triple.} & 13 00 56.3&-76 54 02&         &   1\farcs43$\pm$0\farcs03 &5\fdg9$\pm$1\fdg0&1.98&K1 &12.5 &W
\\ 
&13 04 22.3&-76 50 08&Hn 22/23  &   6\farcs11$\pm$0\farcs05 &227\fdg8$\pm$1\fdg0&1.15 &M1/M2& 13.5+12.2(R)     &\\
&13 04 55.7&-77 39 49&Hn 24  &   1\farcs64$\pm$0\farcs03 &183\fdg6$\pm$1\fdg0&2.44 &     &13.3(R) &\\ \hline
RXJ1517.1-3434 & 15 17 10.7&-34 34 15&   CD-34 10292B&   0\farcs72$\pm$0\farcs04 &223\fdg1$\pm$4\fdg0&2.19&K1&
11.1&W\\
RXJ1525.2-3845 & 15 25 17.0&-38 45 26&HD 137059 &   1\farcs05$\pm$0\farcs02 &217\fdg7$\pm$0\fdg4&0.16&G5V & 
9.3+9.6 &W\\
RXJ1528.7-3117 & 15 28 44.0&-31 17 39&HD 137727  &   2\farcs20$\pm$0\farcs02 &183\fdg7$\pm$0\fdg1&0.65 &K0  & 
9.4+9.9 &W\\
RXJ1530.4-3218 & 15 30 26.3&-32 18 12&HD 138009   &   1\farcs55$\pm$0\farcs02 &25\fdg8$\pm$0\fdg1 &0.14&G6V & 
9.3+9.5 &W\\
RXJ1535.8-2959 &15 35 48.3&-29 58 54&                &   0\farcs90$\pm$0\farcs01 &71\fdg8$\pm$0\fdg4 &0.04&M4&15.3&C\\
RXJ1536.5-3246 &15 36 33.7&-32 46 10&  &2\farcs38$\pm$0\farcs02 &134\fdg6$\pm$0\fdg1&0.36&M3&14.7&W+\\
RXJ1537.0-3136AB & 15 37 01.9&-31 36 37&CD-31 12102 &   5\farcs91$\pm$0\farcs01 &287\fdg4$\pm$0\fdg1&1.37& G5 &  
9.4 &W\\
RXJ1537.0-3136BC & 15 37 01.5&-31 36 35   &CD-31 12102B&   1\farcs41$\pm$0\farcs01 &132\fdg0$\pm$0\fdg1&0.18&K7 & 12.8+12.9 &W\\
RXJ1540.2-3223 & 15 40 16.3 &-32 23 18 &                  &   1\farcs01$\pm$0\farcs01 &61\fdg6$\pm$1\fdg0&1.54&M3&14.4&W\\
RXJ1544.0-3311 & 15 43 39.6&-33 11 24 &   &   1\farcs25$\pm$0\farcs03 &198\fdg2$\pm$0\fdg6&3.28&G8&11.2&W\\
RXJ1545.2-3417\footnote{The binary nature of this star was first
discussed by Brandner (1992) and Reipurth \& Zinnecker (1993).}& 15 45 12.7&-34 17 30    &  HT Lup, Sz 68   &   2\farcs80$\pm$0\farcs02 &297\fdg3$\pm$0\fdg1&3.41&  K0  
& 
10.2 &W+ \\
RXJ1545.7-3020\footnote{primary saturated on the 1$\mu$m image, therefore $\Delta$Z is only a lower limit.} & 15 45 47.6&-30 20 56&HD 140637      &   0\farcs67$\pm$0\farcs03 &348\fdg9$\pm$0\fdg1&$>$2.0&  K3 & 
9.3 &W\\
RXJ1551.4-3131 &15 51 26.8&-31 30 59&                             &   0\farcs82$\pm$0\farcs03 &270\fdg5$\pm$1\fdg0&1.91&  M1  & 
13.5 &W\\
RXJ1552.5-3224 &15 52 30.0&-32 24 12&                &   2\farcs59$\pm$0\farcs01 &261\fdg4$\pm$0\fdg1&0.08&M2&14.9 &W+ \\
RXJ1554.0-2920 &15 54 03.6&-29 20 15&                 &   1\farcs39$\pm$0\farcs01 &75\fdg3$\pm$0\fdg3 &1.34&M0 &12.4 &W\\
RXJ1554.9-2347\footnote{Ghez et al.\ (1993) identified the companion using 2.2$\mu$m
speckle imaging. The values reported there are: sep.\ $0\farcs80\pm0\farcs01$, 
PA $229^\circ  \pm 1^\circ$, and $\Delta$K=2.1.} & 15 54 59.9&-23 47 18&HD 142361 &   0\farcs73$\pm$0\farcs03 &235\fdg7$\pm$3\fdg0 &2.13&G3V&  
9.1 &W \\
RXJ1555.6-3200 &15 55 37.0&-31 59 58&   &   2\farcs95$\pm$0\farcs02 &301\fdg3$\pm$1\fdg0 &5.09&M2&13.8&W\\
 &15 58 07.3&-41 51 48&Sz 128  &   0\farcs60$\pm$0\farcs01 &288\fdg8$\pm$4\fdg9 &0.87&M1&13.5&C\\
RXJ1559.2-2606 & 15 58 53.8&-26 07 21 &  HD 143018 B    &   3\farcs03$\pm$0\farcs02 &329\fdg6$\pm$1\fdg0 &1.72&K2& 
12.3 &W\\
RXJ1605.6-2004 &16 05 42.7&-20 04 15&ScoPMS 029            &   0\farcs69$\pm$0\farcs03 &355\fdg0$\pm$1\fdg0 &0.55&M1&14.3&W+\\
\hline
\end{tabular}
notes:\\
$\Delta$Z: brightness difference between companion and primary at 1$\mu$m.\\
Spectral types and m$_V$ are from Alcal\'a et al.\ (1995) and Kunkel (1995). 
If two values
for m$_V$ are given, they correspond to the brightness of the primary and 
secondary, respectively (Worley \& Douglass 1984). An `R' indicates that
we give the R rather than the V magnitude.\\
E$_{H\alpha}$: H$\alpha$ equivalent width: W (E$_{H\alpha} <$ 0.5nm),
W+ (0.5nm $<$ E$_{H\alpha} <$ 1.0nm), C (E$_{H\alpha}>$ 1.0nm). ''W``
includes stars with H$\alpha$ in absorption (i.e. negative E$_{H\alpha}$) 
\end{minipage}
\end{center}
\end{table*}

In order to achieve a high duty cycle we applied a kind of ``snap shot'' mode.
Only part of the whole CCD,
corresponding to an $1\farcm5 \times 1\farcm5$ field, was read out.
The telescope was already pre-set to the next
programme star during the read-out procedure.
No guide stars were acquired. In each of the two observing nights in 1994
we obtained about 80
scientific exposures within 3 hours each, including the
time spent on image analysis and refocusing the telescope.
The overall efficiency, i.e. the time the CCD shutter
was open in comparison to the total telescope time used, was 20\%. This
value may seem quite low. However, using speckle techniques or
adaptive optics measurements would require a lot more overhead time
without allowing us to survey the fainter stars.
Thus this mode of observation does indeed make relatively efficient use
of telescope time.

In all nights we found the point spread function (PSF) -- 
even after doing image analysis -- to be
slightly elongated. The mean absolute elongation, i.e. the difference
between the FWHM
of the PSF in the direction of elongation and perpendicular to that direction, 
was about 0\farcs13 -- independent of the seeing. See Table \ref{obs} and
Fig.\ \ref{see_sco} for more information on the seeing values.

The data reduction was done following the procedures described by
Brandner (1992, 1993). 
PSFs and close binaries were analyzed by least-squares methods using
GaussFit\footnote{GaussFit is available via anonymous ftp from 
clyde.as.utexas.edu in the directory /pub/gaussfit.}, a `package for 
least-squares-fits and robust estimation' (Jefferys 1991). 
For all detected binary and multiple systems we measured separation, position
angle (PA), 
and brightness difference ($\Delta$Z) between companion and primary 
(cf. Table \ref{cross}).
Simulations showed that the typical relative errors in the separation 
are less than
2\% and in the brightness ratio are less than 5\%. This accuracy is sufficient
for our study.

As is evident from Fig. \ref{see_sco} for 85\% of the exposures
in Scorpius and Lupus, the seeing was 0\farcs65 or less. Because of the 
image elongations
we could however have missed some binaries with a separation larger than 
0\farcs65,
if -- by accident -- the PA between primary and secondary
should coincide with the
direction of elongation. Given a mean absolute elongation of 0\farcs13
we should have resolved all binaries with separations larger
than 0\farcs80, i.e., 120AU at a distance of 150pc.
This value defines our completeness limit.

For the observations of TTS in Chamaeleon in 1994 May 17, the data
set can be split up in two parts: during the first part (one third
of the whole sample) the median value for the seeing (FWHM) was
0\farcs93. After doing image analysis, we obtained a median seeing value of
0\farcs73 for the rest of our sample.
Thus for the first part of these observations we cannot claim
completeness in binary detection down to 0\farcs8 and very likely
have missed some of the closer binaries.
However, in the second part, we should have detected almost all binaries
with separations down to 0\farcs8.

In 1995 February 6 and 7 the seeing was between 0\farcs60 and 1\farcs00.
If the seeing on individual exposures was worse than 0\farcs80, the
stars were reobserved. Again the survey should be complete for binary
separations down to 0\farcs80.
\subsection{Near-infrared Imaging (MPIA 2.2m)}

In 1994 June 17 we observed 13 of the multiple systems in the Scorpius region
with IRAC2B in the J, H, and K band at the MPIA 2.2m telescope on La Silla.
IRAC2B is equipped with a Rockwell NICMOS-3 array. With lens A,
the image scale of IRAC2B is 0\farcs153 and the field size is
$39'' \times 39''$.
Each star was observed twice on different positions on the array.
Thus the second frame could be used for the subtraction of sky
from the first frame and vice versa.

\begin{table}
\begin{center}
\caption{\label{jhk}Near-infrared photometry of selected multiple systems 
in Scorpius and Lupus.}
\begin{tabular}{lrcc}
ROSAT design.& J &J-H&H-K\\ \hline
RXJ1517.1-3434  &10\fm60$\pm$0\fm01  &0\fm70$\pm$0\fm04  &0\fm22$\pm$0\fm04      
  \\ \hline
RXJ1525.2-3845 A& 8\fm15$\pm$0\fm02  &0\fm33$\pm$0\fm03  &0\fm07$\pm$0\fm02      
  \\
               B& 8\fm31$\pm$0\fm03  &0\fm40$\pm$0\fm04  &0\fm11$\pm$0\fm02      
  \\ \hline
RXJ1528.7-3117 A& 7\fm84$\pm$0\fm03  &0\fm44$\pm$0\fm06  &0\fm12$\pm$0\fm06      
  \\
               B& 8\fm59$\pm$0\fm03  &0\fm34$\pm$0\fm06  &0\fm07$\pm$0\fm06      
  \\ \hline
RXJ1530.4-3218 A& 8\fm14$\pm$0\fm01  &0\fm41$\pm$0\fm06  &0\fm10$\pm$0\fm06      
  \\
               B& 8\fm24$\pm$0\fm02  &0\fm40$\pm$0\fm06  &0\fm09$\pm$0\fm06      
  \\ \hline
RXJ1535.8-2959 A&11\fm06$\pm$0\fm08  &0\fm58$\pm$0\fm11  &0\fm28$\pm$0\fm09      
  \\ 
               B&11\fm10$\pm$0\fm12  &0\fm62$\pm$0\fm15  &0\fm29$\pm$0\fm09        \\ \hline
RXJ1536.5-3246 A&11\fm14$\pm$0\fm03  &0\fm64$\pm$0\fm09  &0\fm24$\pm$0\fm09      
  \\ 
               B&11\fm47$\pm$0\fm04  &0\fm62$\pm$0\fm10  &0\fm25$\pm$0\fm09      
  \\ \hline
RXJ1537.0-3136 A& 8\fm42$\pm$0\fm02  &0\fm42$\pm$0\fm06  &0\fm10$\pm$0\fm06      
  \\ 
               B& 9\fm17$\pm$0\fm02  &0\fm67$\pm$0\fm06  &0\fm23$\pm$0\fm06      
  \\ 
               C& 9\fm52$\pm$0\fm03  &0\fm69$\pm$0\fm06  &0\fm18$\pm$0\fm06      
  \\ \hline
RXJ1544.0-3311 A& 9\fm12$\pm$0\fm05  &0\fm51$\pm$0\fm07  &0\fm14$\pm$0\fm07      
  \\
               B&11\fm62$\pm$0\fm25  &0\fm53$\pm$0\fm26  &0\fm18$\pm$0\fm10      
  \\ \hline
RXJ1545.2-3417 A& 7\fm67$\pm$0\fm03  &0\fm71$\pm$0\fm08  &0\fm34$\pm$0\fm09      
  \\ 
               B&10\fm58$\pm$0\fm07  &0\fm80$\pm$0\fm10  &0\fm49$\pm$0\fm09      
  \\ \hline
RXJ1552.5-3224 A&11\fm91$\pm$0\fm04  &0\fm71$\pm$0\fm09  &0\fm26$\pm$0\fm09      
  \\
               B&11\fm97$\pm$0\fm04  &0\fm69$\pm$0\fm09  &0\fm28$\pm$0\fm09      
  \\ \hline
RXJ1554.0-2920 A& 9\fm86$\pm$0\fm03  &0\fm70$\pm$0\fm07  &0\fm19$\pm$0\fm08      
  \\ 
               B&11\fm19$\pm$0\fm03  &0\fm71$\pm$0\fm08  &0\fm21$\pm$0\fm08      
  \\ \hline
RXJ1554.9-2347  & 7\fm56$\pm$0\fm04  &0\fm39$\pm$0\fm11  &0\fm11$\pm$0\fm11      
  \\ \hline
RXJ1605.6-2004  &10\fm22$\pm$0\fm04  &0\fm76$\pm$0\fm08  &0\fm28$\pm$0\fm09      
  \\ \hline
\end{tabular}
\end{center}
RXJ1517.1-3434, RXJ1554.9-2347, and RXJ1605.6-2004 were marginally resolved
on the infrared images, but we could not derive reliable photometry
for the components.
\end{table}

Separation, PA, and relative photometry for the components of each
multiple system were determined in the same way as described above for the
1$\mu$m data. Stars out of the ESO list of infrared standard stars
(Bouchet 1994, priv. communication) were observed to allow for an absolute
flux calibration.
The photometric reduction was done within the IRAF/DIGIPHOT package.
J magnitudes and J-H and H-K colours are listed in Table \ref{jhk}.

A comparison of our photometric results with measurements obtained
with the infrared photometer at the ESO 1m (Kunkel 1995) yields
a difference of 0\fm04 in J and H, and 0\fm03 in K, in the sense
that the stars appear to be brighter in our measurements. No sign for
variability in any of the 13 sources -- apart from the systematic
differences -- could be found.

The seeing varied between 0\farcs8 and 1\farcs3 with a median
value around 1\farcs0.

\section{Selection of an unbiased sample of PMS binaries}

\subsection{Bias induced through X-ray selection}

Despite the homogeneous sky coverage of the RASS, the effective exposure times 
(and hence limiting X-ray luminosities) of the X-ray selected T Tauri stars 
are varying.
In Scorpius the RASS exposure times were about 500 sec, whereas in 
Chamaeleon exposure times were 2.5 times longer.
Finally, the pointed ROSAT
PSPC observations of the dark cloud Chamaeleon I had exposure times of 
2 $\times$ 6 ksec (cf.\ Feigelson et al.\ 1993) and 34 ksec (cf. Alcal\'a 1994,
Zinnecker et al.\ 1996, in preparation), respectively. 

The high sensitivity limit of the RASS and the fact that ROSAT-unresolved 
binaries are statistically brighter X-ray sources than single stars
induces a detection bias. While all single and binary stars with either
one component exceeding the limiting X-ray luminosity L$_x^{limit}$ have been 
detected, additional binaries with component X-ray luminosity L$_x$ just 
below, but combined L$_x$ above the cut-off would cause an overestimate of the 
actual binary frequency. 

\begin{figure}
\centerline{\psfig{figure=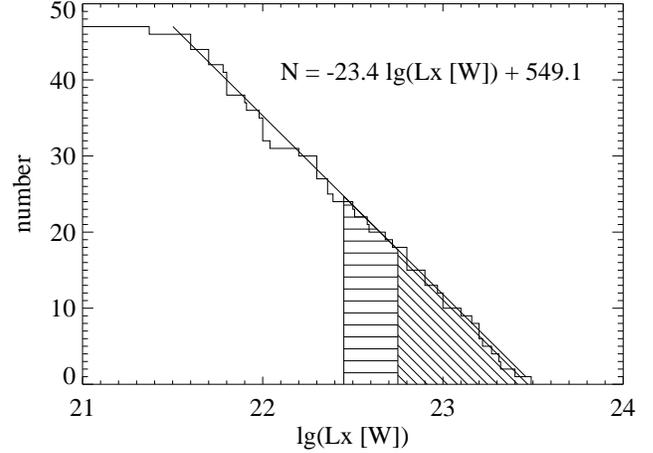,width=9.5cm}}
\caption{\label{lx}. Integrated X-ray luminosity function for 47 
H$\alpha$ selected T Tauri stars associated with the dark cloud Chamaeleon I. 
The diagonal
hatching marks the completeness limited of the RASS in Chamaeleon
for single stars (lg L$_x^{limit}$ [W] $\ge$ 22.75), the horizontal hatched
area indicates the possible bias through unresolved binaries with component
luminosities below the limiting X-ray luminosity, but combined luminosities
exceeding this limit. A linear fit to the integrated X-ray luminosity function
is also plotted.}
\end{figure}

Out of 92 H$\alpha$ detected TTS associated with the dark cloud Chamaeleon I 
(Schwartz 1989, Hartigan 1993), 47 were detected in the pointed ROSAT 
observations.
In Fig.\ \ref{lx} we show the integrated X-ray luminosity function of these 
TTS. X-ray luminosities are from Feigelson et al.\ (1993) and Alcal\'a 
(1994).
As has been shown already by Bouvier (1990) for TTS in Taurus and has been
confirmed by Alcal\'a (1994) for TTS in Chamaeleon, there is
no apparent correlation or anticorrelation between X-ray and H$\alpha$
luminosity. Hence, the H$\alpha$ selection criterion should not induce any
additional bias and the integrated X-ray luminosity function should
be a good representation for the overall TTS population.

A linear fit yields
$$\frac{\mbox{dN}}{\mbox{dL$_x$}} \propto \mbox{L$_x$}^{-1} \qquad
\mbox{for}\quad 10^{21.5}\mbox{W} \le \mbox{L$_x$} \le 10^{23.5}\mbox{W,}$$
i.e.\ in equal bins of L$_x$ the number of stars varies with
$\mbox{L$_x$}^{-1}$. The distribution function diverges for
$\mbox{L$_x$} \rightarrow
0$, but we are anyway only interested in binaries with at least one binary
component exceeding $0.5 \mbox{L$_x^{limit}$}$.

By assuming that the stars detected 
in the
RASS follow a X-ray luminosity function similar to the one obtained
for the dark cloud Chamaeleon I, we can estimate the number 
of additional 
binaries in our survey and correct for this detection bias. 

If we have a sample of N RASS sources, only $\mbox{N} - \Delta\mbox{N}$ form 
an unbiased subsample, the remaining $\Delta$N are binaries with the combined 
luminosity exceeding the limiting X-ray sensitivity.
Without the use of a priori knowledge on the pairing preferences of TTS, we
can estimate the maximal $\Delta\mbox{N}$, assuming that all binaries
with one component's X-ray luminosity exceeding $0.5 \mbox{L$_x^{limit}$}$ 
have a combined  X-ray luminosity exceeding $\mbox{L$_x^{limit}$}$.
The completeness limit of the RASS in Chamaeleon is
lg L$_x^{limit}$ [W] = 22.75 (Alcal\'a et al.\ 1995).
\begin{figure}
\centerline{\psfig{figure=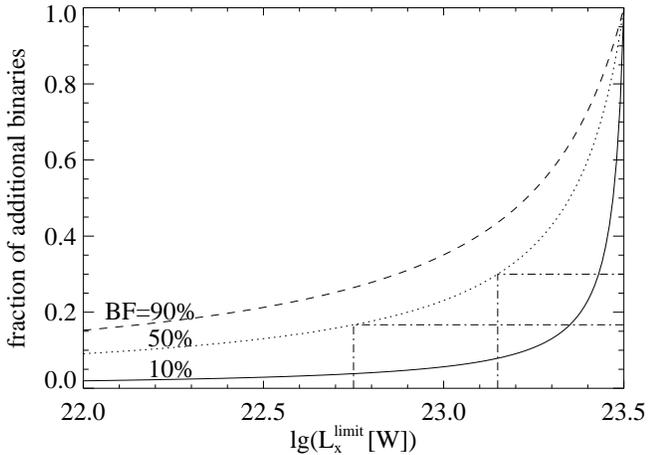,width=9.5cm}}
\caption{\label{blx} Maximal fraction of additional binaries among RASS sources
with component X-ray luminosity below, but combined X-ray luminosity
above the sensitivity limit 
as a function of limiting RASS X-ray sensitivity and for binary frequencies (BF)
of 10\%, 50\%, and 90\%.}
\end{figure}
In Fig. \ref{blx} we have plotted $$\Delta\mbox{N/N} =
\frac{
\int_{0.5 \mbox{L$_x^{limit}$}}^{\mbox{L$_x^{limit}$}} 
{\mbox{L}_x}^{-1}\mbox{dL$_x \times$ BF}
}{
\int_{\mbox{L$_x^{limit}$}}^{10^{23.5}\mbox{W}} 
{\mbox{L}_x}^{-1}\mbox{dL$_x$} +
\int_{0.5 \mbox{L$_x^{limit}$}}^{\mbox{L$_x^{limit}$}} 
{\mbox{L}_x}^{-1}\mbox{dL$_x \times$ BF}
} $$ 
as a function of
limiting X-ray sensitivity
% L$_x^{limit}$ 
and binary frequency (BF).

If we assume a binary frequency of 50\%, we get $\Delta\mbox{N}$/N = 1/6
for the RASS sources in Chamaeleon. The extra number of binaries
$\Delta$N is 95/6 = 16 and the total number of binaries in our biased sample 
is 16+79/2 = 55.5.

This can be compared to the actual number of detected binaries (6 systems),
yielding a ``success rate'' of 6/55, i.e. 11\%. Applying the same
success rate to the extra number of binaries due to detection bias,
we find 16 $\times$ 0.11 $\approx$ 1.7 systems in excess of what we
would have found in an unbiased survey. Therefore the corrected
binary frequency is $\frac{6-1.7}{95-15.8} = \frac{4.3}{79.2} \approx 5.4\%$
(to be compared with 6.3\% in the biased survey).

In Scorpius  the X-ray luminosity completeness limit is somewhat higher 
because of the shorter RASS exposure times (lg L$_x$ [W] $\ge$ 23.15,
Kunkel 1995). 
From Fig.\ \ref{blx} we get $\Delta\mbox{N}$/N = 0.3, i.e. 22.2 additional 
binaries among the 74 TTS.
Following
the same reasoning as above, we derive a ``true'' binary frequency of 
7.0/51.8 = 13.5\% (instead of 13/74 = 17.6\% without correction).

Overall binary frequencies of 10\% or 90\% would induce smaller errors as
i) for a large sample and a small binary frequency a few additional 
binaries would not make that much of a difference and ii) for a high binary 
frequency the relative number of binaries in the biased and unbiased 
sample would not change that much.

The ``debiased'' binary frequencies derived in this section 
present lower limits (since we assumed that all binaries with
one component's X-ray luminosity exceeding 0.5 L$_x^{limit}$ would
be detectable)
and have of course still to be corrected for chance projections and 
incompleteness (cf.\ section 7.2).

\subsection{Notes on individual programme stars}

In the following we briefly discuss some of the surveyed objects
(see also the footnotes to Table \ref{cross}):

\begin{description}

\item[RXJ1301.0-7654] This star is also a spectroscopic binary (Covino
et al.\ 1995) and thus very likely forms a hierarchical triple
system.

\item[RXJ1537.0-3136] The only visual triple system in the whole survey. While
magnitudes and separations measured by us agree well with the values
listed in the Washington Double star catalogue (Worley \& Douglass 1984), 
the PAs do not match at all. From spectral types
and V magnitudes we get a ZAMS distance of 70--80pc for these stars,
hence they are too distant in order to explain the discrepancy
in PA by orbital motion.

\item[RXJ1540.2-3223] The identification as a TTS is uncertain. Li\,I
is only marginally resolved, H$\alpha$ is in absorption (Kunkel 1995).

\item[RXJ1554.9-2347] This star has already been identified by Walter
et al.\ (1994) on the basis of EINSTEIN data as a TTS (ScoPMS 005). 
Ghez et al.\ (1993) reported a companion (see footnotes to
Table \ref{cross}).

\item[RXJ1602.1-2241] 
Mathieu et al.\ (1988) found this star to be an SB1 (P=2.4d, e=0.02). 
Ghez et al.\ (1993) identified  a tertiary (sep. 0\farcs288, PA 347$^\circ$,
$\Delta$K = 0\fm83).
This triple system (ScoPMS 023 in Walter et al.\ 1994) is unresolved in our survey.

\item[RXJ1605.6-2004] This star has been identified by Walter et al.\
(1994) on the basis of EINSTEIN data as a TTS (ScoPMS 029).

\end{description}

\subsection{Selection criteria for physical pairs: separation}

Before trying to interpret the observations we have to be sure that we
indeed observed physical binaries and not mere chance projections.
Contrary to most CTTS which can be found in or close to molecular clouds,
WTTS are spread out over a much wider area of the sky where no dark
cloud provides an effective ``screen'' against contamination by
background stars, especially giants. Both major regions -- Scorpius/Lupus
as well as Chamaeleon -- lie well above or below the galactic plane,
respectively ($b \approx \pm 15^\circ$). At a galactic latitude
of 15$^\circ$ the stellar density is 500 stars/square degree if we go to 
a limiting photographic
magnitude of 16, 5000 stars/square degree  for 18$^{\rm m}$, and
8000 stars/square degree for 19$^{\rm m}$ (Scheffler \& Els\"asser, 1965).
Unlike Chamaeleon, Scorpius lies in the direction to the
galactic bulge (l = $340^\circ$ -- $0^\circ$). In this direction
the stellar density is even higher and reaches 9000 stars/square degree  for 
18$^{\rm m}$.
Thus, the fainter the companion the more likely it is just a chance
projection. The faintest secondary in our list is about 18$^{\rm m}$ in V.
The probability to find a star down to 18$^{\rm m}$ within an
aperture of 3$''$ radius is 1\% to 2\%. Hence we would expect 3
false detections among the about 200 stars surveyed. RXJ1555.6-3200 (m$_{V\rm prim}$=13\fm8,
$\Delta$Z = 5\fm1, sep = 3\farcs0) is a good candidate for such a chance
projection.

If we allowed an aperture radius of 12$''$ and included all stars
down to a limiting magnitude of 18 in V, the probability to find a star
within this aperture would already be 20\% to 40\%! Spectroscopic
follow-up observations of faint secondaries to CTTS in Chamaeleon
from the list of Brandner (1992) showed that almost all of them
are indeed background giants (Brandner 1995).
In order to minimize false ``detections'' we only consider binaries
with a projected separation of 3$''$ or less as sure detections.
In the range of separations from 3$''$ to 12$''$ we have a number of
binary candidates. For these objects photometric, spectroscopic, and astrometric
follow-up studies are necessary to tell physical binaries from mere
chance projections.

\subsection{Near-infrared checks of physical companions}

\begin{figure}
\centerline{\psfig{figure=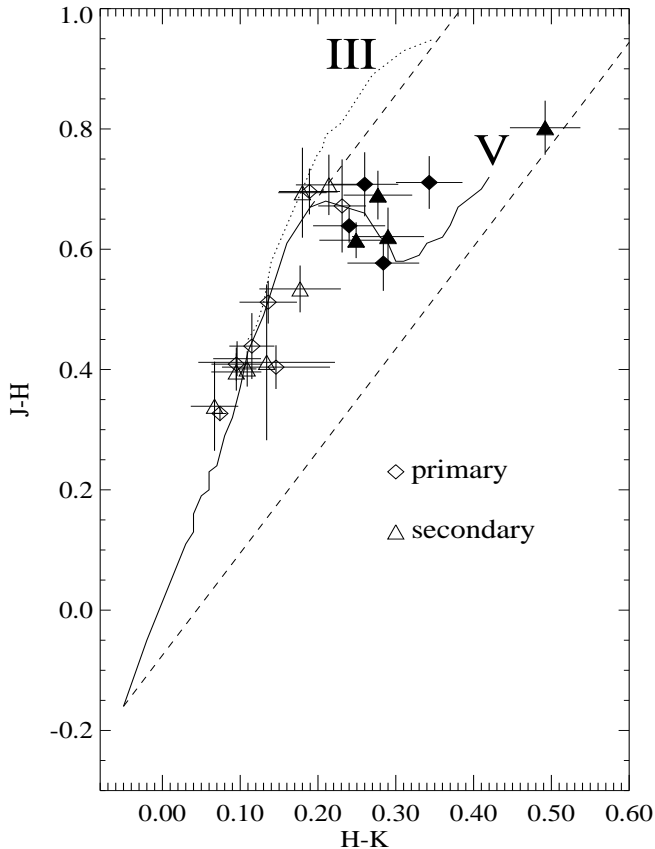,height=12cm,width=9.5cm}}
\caption{\label{sco_jhk}Near-infrared colour-colour diagram of
the components of binary systems in the Scorpius and Lupus region.
The solid line marks the main sequence ({\em V}), the dotted line giant 
stars ({\em III}).
The area between the two
dashed lines indicates the region in which reddened main sequence stars
would fall. Filled symbols mark systems with strong H$\alpha$ emission
(E$_{H\alpha} > $ 0.5nm). The photometric uncertainties are indicated.}
\end{figure}

The near-infrared colour-colour diagram \jh\ vs. \hk\ is shown in Fig.\
\ref{sco_jhk}. For comparison, the location of main
sequence and giant stars is also indicated. With the exception of
RXJ1545.2-3417 (Sz 68), a classical T Tauri star  (E$_{H\alpha} \approx
0.6$nm) associated with a reflection nebulosity (e.g., Reipurth \&
Zinnecker 1993) and its companion, none of the sources, primaries or
secondaries, show any intrinsic IR-excess or any significant foreground
extinction. Within the uncertainties, all
stars have colours of dwarf stars. The lack of extinction indicates that
all of the identified companions are not background stars
but members of the association. The fact that both Sz 68 and its
companion show an IR-excess makes it very likely that they both belong
to the nebulosity and hence form a physical binary.

Systems with rather strong H$\alpha$ emission (E$_{H\alpha} >$ 0.5nm) are marked
by filled symbols. With the exception of Sz 68, they are of spectral type
M2 or later. 
Thus, similar to the main sequence where we can find
chromospherically active dMe stars, very-late type stars among 
X-ray selected T Tauri stars show strong activity.

Within the photometric uncertainties the NIR-colours for most primaries in general 
agree quite well with their spectral type and are consistent with luminosity
class V. The two exceptions are RXJ1528.7-3117 (SpT K0), whose
NIR-colours would suggest a slightly later spectral type for the
primary (K1) and an earlier spectral type for the secondary (G8), and
RXJ1544.0-3311 (SpT G8), whose NIR-colours suggest a spectral type
of K2/K3 for the primary. 

\subsection{Common proper motion pairs}

\begin{table*}
\caption{\label{astrom}Astrometric data}
\begin{center}
\begin{minipage}{160mm}
\begin{tabular}{lllllc|rr}
ROSAT design. & IDS design. & epoch & separation & PA & reference & $\mu_\alpha$ in s&$\mu_\delta$ in $''$\\
\hline
RXJ1039.5-7538&IDS 10374-7507 A & 1931&4\farcs9& 347 & WDS&& \\
              & & 1992.1&5\farcs04$\pm$0\farcs05& 351\fdg2$\pm$1\fdg0 & B&& \\ \hline
RXJ1108.1-7742& & 1992.1&0\farcs69$\pm$0\farcs01& 178\fdg7$\pm$3\fdg2 & B&& \\
              & & 1994.5&0\farcs72$\pm$0\farcs03& 178\fdg6$\pm$1\fdg0 &  BAKMZ&& \\ \hline \hline
RXJ1525.2-3845 &IDS 15189-3823 BC & 1897 & 1\farcs3 & 247$^\circ$ & WDS&-0.0032&-0.033 \\
              & & 1980 & 1\farcs1 & 224$^\circ$ & WDS&& \\
              & & 1994.3 & 1\farcs05$\pm$0\farcs02& 217\fdg7$\pm$0\fdg4& BAKMZ&& \\ \hline
RXJ1528.7-3117 &IDS 15226-3057 AB & 1913 & 2\farcs2 & 179$^\circ$ & WDS&0.0012&0.047 \\
              & & 1994.3& 2\farcs20$\pm$0\farcs02& 183\fdg7$\pm$0\fdg1 & BAKMZ&& \\ \hline
RXJ1530.4-3218 &IDS 15242-3158 B & 1928 & 1\farcs3 & 40$^\circ$ & WDS&-0.0022&-0.030 \\
              & & 1938 & 1\farcs3 & 37$^\circ$ & HIC&& \\
              & & 1961 &          & 31$^\circ$ & WDS&& \\
              & & 1994.3& 1\farcs55$\pm$0\farcs02& 25\fdg8$\pm$0\fdg1 & BAKMZ&& \\ \hline
RXJ1537.0-3136 A--BC&IDS 15308-3117 A & 1913 & 4\farcs5& 32$^\circ$ & WDS&-0.0021&-0.033 \\
               & & 1961 &         & 81$^\circ$ & WDS&& \\
               & & 1994.3&5\farcs2&283$^\circ$\footnote{sep.\ and PA are between
A and the ``center of light'' of BC.} & BAKMZ&& \\ \hline
RXJ1537.0-3136 BC&IDS 15308-3117 BC &1934 & 1\farcs4 & 23$^\circ$ & WDS&& \\
               & &1945 &          & 7$^\circ$  & WDS&& \\
              & &1994.3& 1\farcs41$\pm$0\farcs02&132\fdg0$\pm$0\fdg2\footnote{a PA of 312$^\circ$ is
also possible, since C is brighter than B in the visual (Kunkel 1995). This,
however, still cannot explain the large discrepancies between the PA mesasured
by us and the PAs given in WDS. PAs and separation observed by Kunkel
in 1993.5 are in agreement with the values determined by us. Hence there is
no indication for a fast change in the relative positions of the stars.} & BAKMZ&& \\ \hline
RXJ1545.2-3417 & & 1991.3 & 2\farcs6 & 298\fdg0$\pm$3\fdg0 & B, RZ&& \\
              & & 1994.3& 2\farcs80$\pm$0\farcs02&297\fdg3$\pm$0\fdg1& BAKMZ&& \\ \hline
RXJ1554.9-2347& &1990.5& 0\farcs80$\pm$0\farcs01 &229$^\circ$$\pm$1$^\circ$ &GNM&-0.0020&-0.034  \\
             & &1994.3& 0\farcs73$\pm$0\farcs03 &235\fdg7$\pm$3\fdg0 &BAKMZ&& \\ \hline
\end{tabular}
References: B stands for Brandner (1992), BAKMZ for this work, GNM
for Ghez et al.\ (1993), HIC for Hipparcos Input Catalogue (Turon et al.\ 1993),
RZ for Reipurth \& Zinnecker (1993),
 and WDS for the Washington Double Star Catalogue
(Worley \& Douglass 1984). Proper motions are from Bastian \& R\"oser
(1993). The IDS designations are out of the WDS.\\
Note that position angles are unprecessed (i.e., they are for the mean date of 
observation).
\end{minipage}
\end{center}
\end{table*}

Positions and proper motions for many
stars brighter than 10$^{\rm m}$ in V can be found in the ``Positions and Proper 
Motions star catalogue'' (PPM, Bastian \& R\"oser 1993).
Some of the brighter binaries are already listed in the Washington
Double Star catalogue (WDS, Worley \& Douglass 1984). 
In Table \ref{astrom} we have listed all binaries for which proper motions are
available and/or for which at least two relative astrometric measurements
from different epochs exist.
In the following 
we discuss some of these objects in greater detail:

\begin{description}
\item[RXJ1525.2-3845] This star is the binary with the longest
time base. However, only 1/12th of a whole orbit has been observed yet.
The star has moved about 3\farcs2 south and 3\farcs6 
west in 97 yrs. The similarity between the separations and position angles
measured in 1897, 1980, and 1994 implies that both components share the
same proper motion. Hence, we attribute the changes in separation and PA
solely to orbital motion: The PA has changed by 23$^\circ$ in 83yrs
from 1897 to 1980 and by 6$^\circ$ in the following 14yrs. At the same
time the projected separation has shrunk from 1\farcs3 to 1\farcs05.
This would imply a system with an eccentricity larger than 0.2
approaching its periastron.

\item[RXJ1528.7-3117] This star has moved about 3\farcs8 north and 1\farcs2
east in 81 yrs. The constancy of the separation (2\farcs2)
measured in 1913 and 1994 implies that both components share the
same proper motion. The change in PA can be attributed to orbital
motion.

\item[RXJ1530.4-3218] This star has moved about 2\farcs0 south and 1\farcs8
west in 66 yrs. The similarity between the separations and positions angles    
measured in 1928, 1938, 1961, and 1994 implies that both components share the
same proper motion. The change in separation and PA can be attributed
to orbital motion: The system's PA has changed by 9$^\circ$ in 33yrs from 1928
to 1961 and by 5$^\circ$ in the following 33 yrs. At the same time the
separation has grown from 1\farcs3 to 1\farcs55. This would imply a
system with an eccentricity larger than 0.2 approaching its apastron.

\item[RXJ1537.0-3136] The only visual triple system observed in our survey. The
primary has moved about 2\farcs7 south and 2\farcs2 west in 81 yrs.
We note that the discrepancy in PA between our measurement and the
values given in the WDS catalog cannot be
explained by proper motion or orbital motion. 
On the other hand, separations and V magnitudes are
in good agreement with our measurements. Observations obtained by Kunkel
(1995) in 1993.5 with the CCD camera at the Dutch telescope on La Silla confirm our measurements.

\item[RXJ1554.9-2347] This star has moved about 0\farcs14 south and 0\farcs11
west in 4 yrs from 1990 to 1994. 
If the companion were a background object -- and therefore
had no significant proper motion within 4 yrs -- , we would
expect a separation of 0\farcs63 and a PA of 232$^\circ$ in 1994.
The observed changes in separation are smaller and in PA are larger,
hence we suggest that both components are association members
with similar proper motions, and are very likely gravitationally bound (the
relatively large change in PA could be explained by 
a high eccentricity orbit, i.e. e $>$ 0.5 -- but note the uncertainty
of $\pm 3^\circ$ in our measurements of the PA).
\end{description}

For the remaining objects in Table \ref{astrom} no proper motion measurements
could be found.
Hence for all systems where we know both proper motion and relative 
astrometry of the components over a time base from 4 to 97 yrs we can 
show that they are very likely physical binary systems.

If we make some simplified assumptions, i.e.,
circular orbits perpendicular to the line of sight,
we can -- with the knowledge of the spectral type and luminosity
of the primary -- get a rough estimate of the physical dimensions 
of the orbits and
hence the distance to the binaries. As most primaries described above
are of spectral type G and have a luminosity of $\approx$ 6L$_\odot$, 
we derive a system mass of
$\approx$ 2M$_\odot$.
% RXJ15287-3117 3.725 0.835 RXJ15304-3218 3.73 0.837 RXJ15549-2347 3.76 0.802

For RXJ1525.2-3845, RXJ1528.7-3117, RXJ1530.4-3218, and RXJ1554.9-2347 this leads
to a
distance of 130 pc, 164pc, 129 pc, and 57 pc, respectively.

We note that if the real orbits deviate from the assumption of circular orbits perpendicular to the line of sight this would cause errors in our distance estimate:
 an (in reality) inclined orbit would result in an overestimate of the
distance whereas a high eccentricity orbit with the secondary near
periastron could result in an underestimate
of the system's distance. Furthermore, the short time base of only 4yrs 
and the relatively large uncertainty in the PA 
makes the distance estimate for RXJ1554.9-2347 very uncertain.

\section{Spatial distribution of binaries}

\begin{sidefigure}
\centerline{\psfig{figure=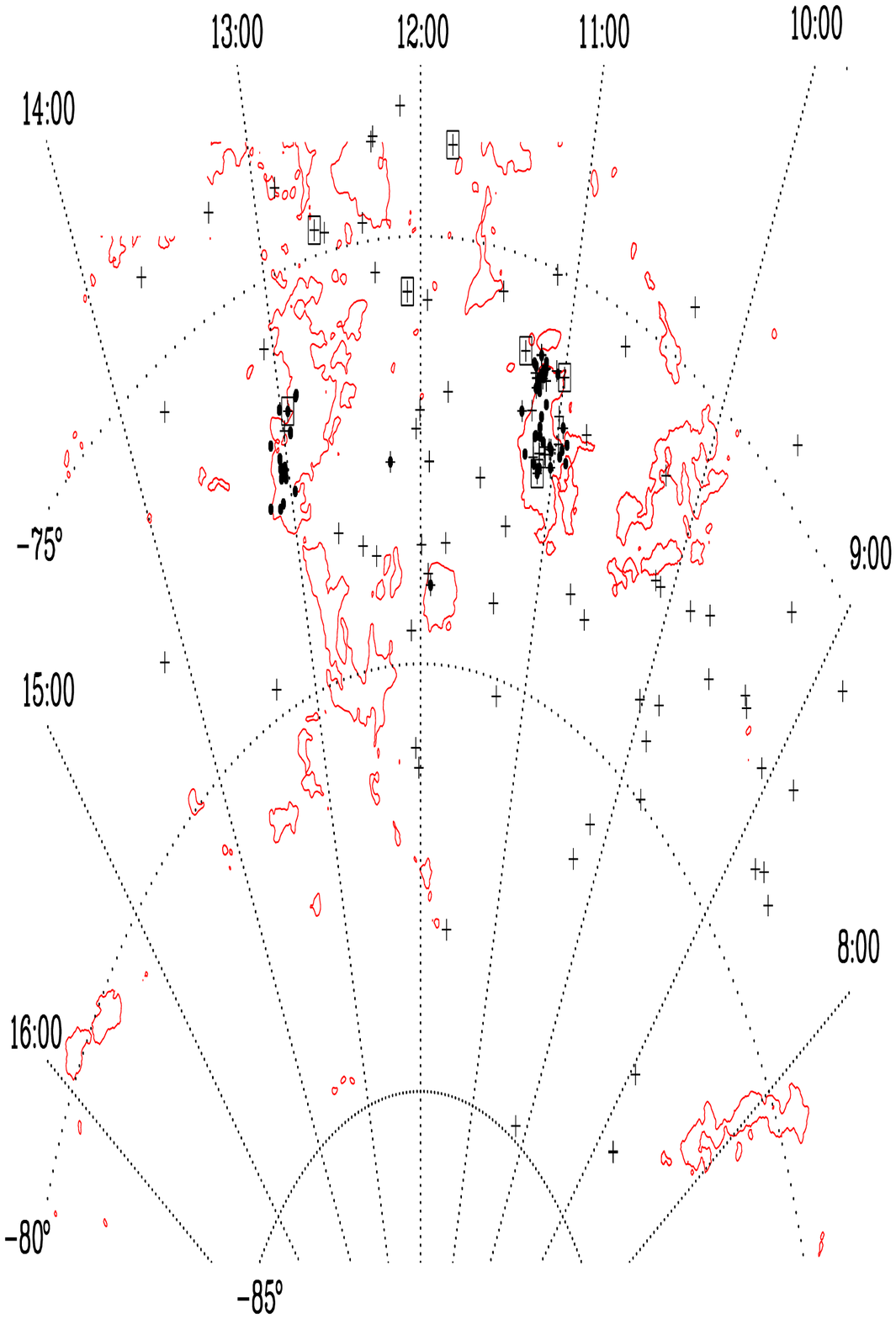,width=26cm,height=16cm}}
\caption{\label{cha_dist}Adapted from Krautter et al.\ (1996, in preparation):
Spatial distribution of CTTS surveyed by
Reipurth \& Zinnecker (1993, filled dots)
and X-ray sources, identified as T Tauri stars by Alcal\'a et al.\ (1995) or 
Huenemoerder et al.\ (1994) and
surveyed by us (plus signs) in the Chamaeleon region.
The IRAS 100 micron contour map is overplotted.
The boxes mark TTS binaries
resolved with SUSI in our current study. Two stars are plotted as CTTS and
WTTS as the strong variability in their H$\alpha$ emission line makes
a classification either as CTTS or WTTS uncertain.}
\end{sidefigure}

Figure \ref{cha_dist} shows the spatial distribution of CTTS and WTTS in
Chamaeleon, adapted from Krautter et al.\ (1996, in prep.).
CTTS are associated with the dark clouds
Chamaeleon I (west) and Chamaeleon II (east). Obviously, WTTS do spread out
over a much wider area than CTTS do (Alcal\'a 1994, Krautter et al.\ 1996 
in prep.).
The binaries among the WTTS, however,
can only be found in a relatively small area defined by the CTTS and
the dark clouds.
This lack of companions among WTTS could be understood, if
almost all WTTS
started as members of (non-hierarchical and thus unstable) triple systems,
where they have been thrown out at a very early stage of PMS evolution
(Harrington 1975).
However,
the WTTS under study are on average more massive than the CTTS (Alcal\'a 1994),
making this scenario very unlikely.
On the other hand, this lack of binaries can at least partly be explained
by the inferior seeing conditions at the time of our survey in Chamaeleon.

\begin{figure*}
\centerline{\psfig{figure=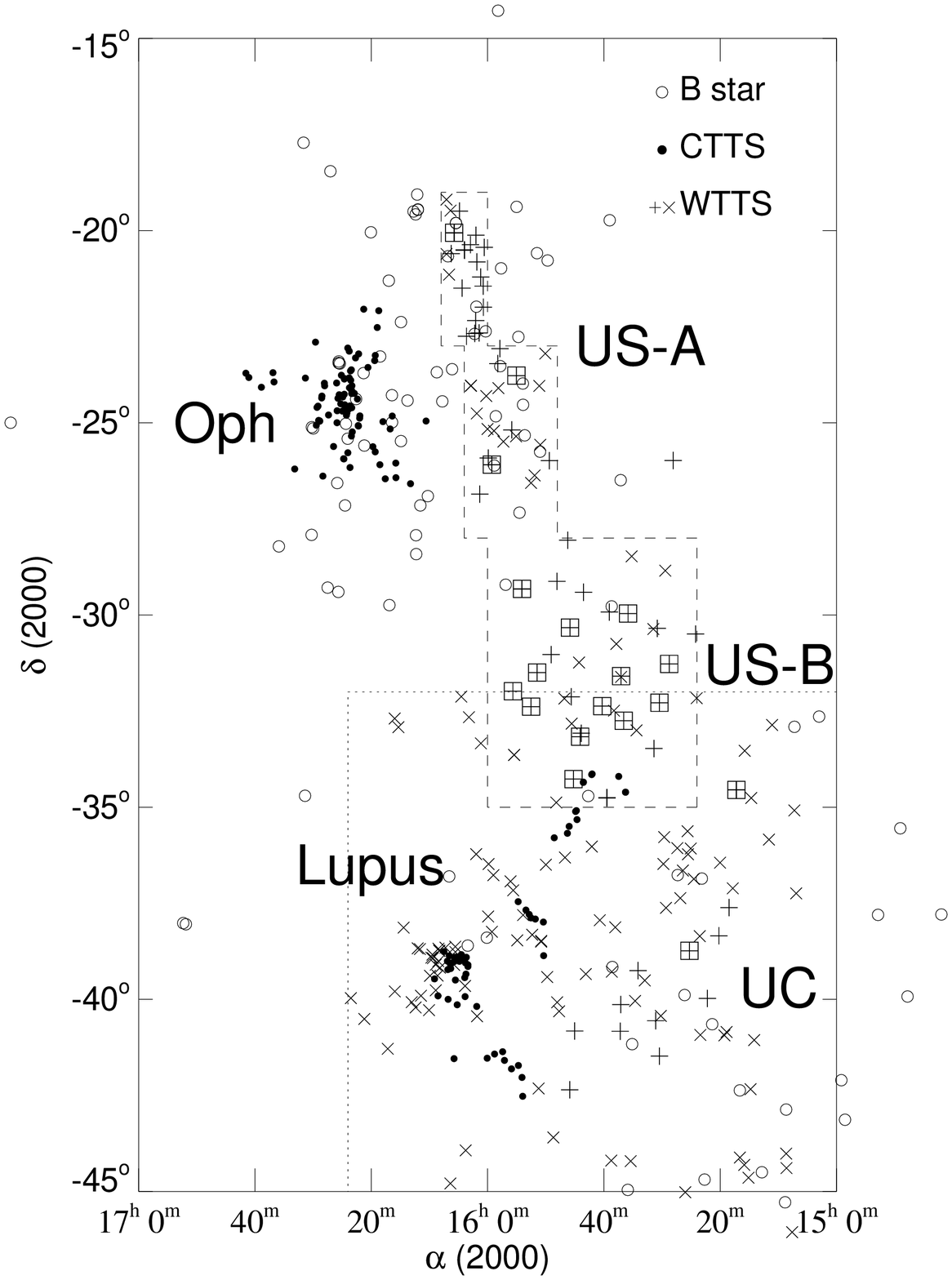,width=16cm,height=21.0cm}}
\caption{\label{sco_bin} Spatial distribution of B stars and T Tauri stars
in the region of Ophiuchus, Scorpius, and Lupus: T Tauri stars identified 
as optical counterparts to ROSAT All Sky Survey sources are marked by a cross or a plus sign
(The latter indicates that these stars have already been searched for
companions using SUSI/NTT).
Classical T Tauri stars associated
with the dark clouds of $\rho$ Ophiuchi and Lupus 1, 2, 3, and 4 are
indicated by filled circles.
The area surveyed by Kunkel (1995)
 for pre-main sequence stars is marked by dashed lines and the northern
and eastern
border of the area surveyed by Wichmann (1994) is indicated by a dotted line.
Open squares indicate binaries resolved with SUSI/NTT in our current
study.}
\end{figure*}

The spatial distribution of B stars that are members of Upper Scorpius and
Upper Centaurus--Lupus 
(de Geus et al.\ 1989) and TTS in
the Ophiuchus--Scorpius--Lupus region is plotted in Fig. \ref{sco_bin}.
TTS identified as optical
counterparts to RASS sources show a rather uniform spatial distribution whereas
CTTS have a clear tendency for clustering. 
To the south, the four distinct groups of CTTS correspond to the 
Lupus T association with the dark clouds Lupus 1 to 4 (from the north to 
the south, Schwartz 1977).
To the north lies the $\rho$ Ophiuchi T association (Wilking et al.\ 1987).

The region originally surveyed by Kunkel (1995) for WTTS is marked by
dashed lines. We further subdivided Upper Scorpius (US) into two distinct 
areas: 

\vspace{2mm}

\noindent
US--A
($\alpha = 16^h0^m$ to $16^h8^m$, $\delta = -19^\circ$ to $-23^\circ$ ) and

\hspace{4mm} ($\alpha = 15^h48^m$ to $16^h5^m$, $\delta = -23^\circ$ to $-28^\circ$ ),

\vspace{2mm}

\noindent
US--B
($\alpha = 15^h25^m$ to $16^h0^m$, $\delta = -28^\circ$ to $-35^\circ$ ).

While US--A lies completely within the Upper Scorpius association, US--B
is located near its periphery at the intersection between Upper Scorpius
and Upper Centaurus--Lupus.
The T Tauri star counterparts to  RASS sources identified by
Wichmann (1994) in the Lupus region  have not yet been
surveyed for companions.

Reipurth \& Zinnecker (1993) investigated the incidence 
of binaries in various southern clouds and found a relation between
binary frequency and the size of the individual clouds.
We now discuss the distribution
of binaries in Scorpius in greater detail.
Among the wide--spread distribution of WTTS there is at least
one ``local surface density enhancement'' 
in the northern part of US--A, where we find 21 TTS and 5 B stars within an
area of $\approx$ 7.5 square degree, i.e. 2.8 TTS/square degree (similar to the Orion
region, cf. Sterzik et al.\ 1995). 
In this particular region almost every  RASS source has a T Tauri star as an 
optical counterpart. Interestingly, among the 17 TTS observed with SUSI/NTT,
only one binary could be detected (RXJ1605.6-2004).
Another small clustering can be found at $\alpha$ = 15$^h$25$^m$ and 
$\delta = -36^\circ$.
Here we have 10 TTS within 2 square degree, i.e. 5 TTS/square degree. These stars
have not been surveyed for multiplicity yet.
A much looser association of TTS is located in US--B, where Wichmann 
(1994) and Kunkel (1995)
identified 36 TTS within 52 square degree, i.e. 0.7 TTS/square degree. Here 13 of the
24 observed TTS show a companion. 

In US--A with its higher spatial density of X-ray selected TTS
($\approx$ 1.5 TTS/square degree) and B stars (0.5 B stars/square degree), 
only a very
low binary frequency (BF, 3/25 = 12\%$\pm$7\%) was found, whereas in US--B
with a moderate spatial density of X-ray selected TTS ($\approx$ 0.7 TTS/square 
degree) and B stars ($<$ 0.1 B stars/square degree)
the BF is higher (13/24 = 54\%$\pm$15\%). If we accept this
difference in BF (12\%$\pm$7\% vs.\ 54\%$\pm$15\%) to be significant -- 
despite the ``threats''  of small number statistics, we have to ask the
question why such a distinction could exist.

One possibility would be that US--A is more distant than US--B, 
so that the difference in BF could be explained
by the difference in spatial resolution. Clearly, the more nearby an
association is, the more easily we can resolve binaries within the region.
The B star population of Upper Scorpius (including the runaway O type star $\zeta$ Oph)
has a mean distance modulus of 6\fm0$\pm$0\fm8 (163pc, see de Geus et 
al. 1989).
The B star population of Upper Centaurus--Lupus, to which the B stars in the south western
corner of Fig. \ref{sco_bin} belong, has a mean distance modulus
of 5\fm8$\pm$0\fm7 (145pc).
The three B stars within US--B have photometric
distances of 100pc, 100pc, and 130pc, respectively (de Geus et al.\ 1989).
Hence, if the TTS stars are physically associated with the B star
population, TTS in US--B are more nearby than those in US--A.
This is in agreement with the finding that the WTTS in US--B show
 a larger Li depletion and hence are presumably older and less luminous
than WTTS in US--A (Kunkel 1995).
However, the difference in distance is too small to fully explain 
the difference in BF.
The distance estimates derived from the relative astrometric measurements
of some of the binaries (cf. previous section) are in good agreement with a 
distance between 130pc and 170pc.

Given the fact that we have a difference in binary frequency in the observed range
of separations, this could either mean that the overall binary frequency
is the same in both regions, but the distribution of separations
among the binaries in the two regions is vastly different.
Or we indeed have fewer binaries in the denser association
and more in the less dense association of TTS. Durisen \& Sterzik (1994)
suggested such a distinction between high-temperature clouds 
(``clustered  star formation'' with lower BF)
and low-temperature clouds (loose associations with higher BF).
Observational evidence for this model comes from HST observations
of the Trapezium cluster (Prosser et al.\ 1994) 
and the various surveys for 
binaries in Taurus (see Mathieu 1994 for a compilation of recent
studies): the Taurus T association has a BF significantly
higher than the main sequence BF, while the BF in the Trapezium cluster
seems to be in agreement with the main sequence BF.
The spatial density of 0.5 B stars/square degree in US--A is far from what 
could be called a cluster, but it is 8 times higher than
in US--B (see Fig. \ref{sco_bin}). 

Reipurth \& Zinnecker (1993) found a weak trend that clouds with large 
populations of young stars may contain fewer binaries. 
Given the wide 
spread distribution of WTTS in our study, and the current lack
of additional information on the kinematics of the individual stars, we cannot 
further subdivide the WTTS into smaller groups in order to investigate
such a trend. 

\section{Distribution of separations and brightness differences}

\begin{figure}
\centerline{\psfig{figure=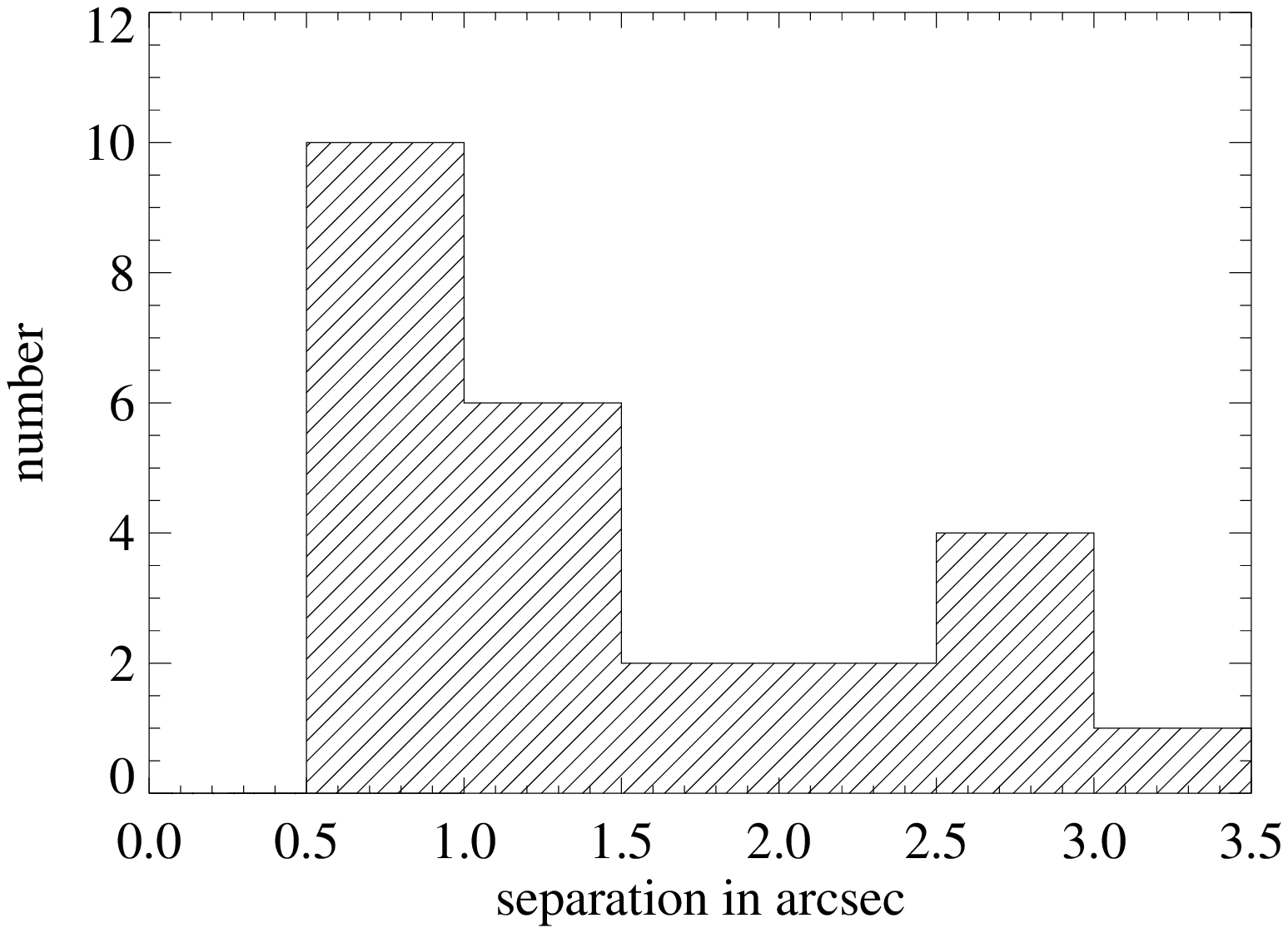,width=9.5cm}}
\caption{\label{bin_sep}Histogram of the distribution of separations
of WTTS binary components in Upper Scorpius and Chamaeleon in bins of
0\farcs5.}
\end{figure}

\begin{figure}
\centerline{\psfig{figure=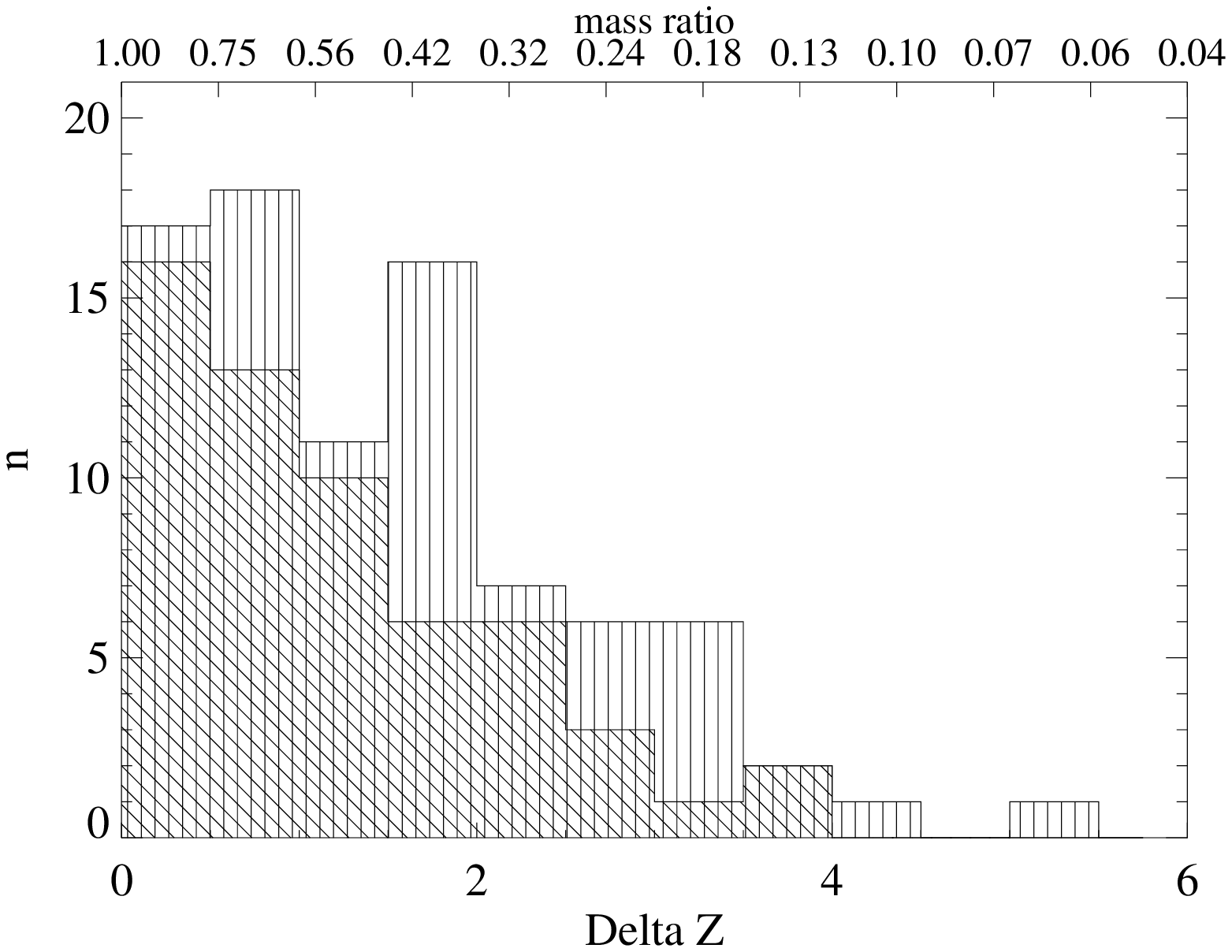,width=9.5cm}}
\caption{\label{deltaz}Histogram of the distribution of brightness
differences between primary and secondary at 1$\mu$m (``Z'') in bins of
0\fm5 for sep.\ $\ge 0\farcs8$ of all TTS from Table 5 for which
measurements were available (vertical hatching). 
For comparison we also show the distribution of brightness differences
of CTTS binaries from Reipurth \& Zinnecker, 1993 (diagonal hatching).  
The upper scale indicates
the corresponding mass ratio between secondary and primary
 assuming purely vertical (T$_{\rm eff}$ = const.) PMS 
evolutionary tracks (see text for more details).}
\end{figure}

Figure \ref{bin_sep} shows the distribution of binary separations
among WTTS in Upper Scorpius and Chamaeleon.
The number of binaries grows with smaller separation. Because
of the uncertainties in the distance estimates of the binaries, however,
we did not try to plot the binary frequency versus the physical separation.

The distribution of brightness difference 
between primary and secondary at $\approx$ 1$\mu$m for TTS binaries 
with separations $\ge$ 0\farcs8 in shown in Fig.\ \ref{deltaz}. Included are 
TTS located in Chamaeleon, Lupus, $\rho$ Ophiuchi, and 
Taurus from the lists of Brandner (1992, 1995), Reipurth \& Zinnecker (1993),
and this paper. Stars with separations smaller than 0\farcs8 have not been
included since they show a clear detection bias towards equal brightness
pairs.

As the maximum contribution of the stellar photosphere to the total spectral 
energy distribution is at about 1$\mu$m (Bertout et at. 1988, Hartigan et al.\ 
1992), the 1$\mu$m observations give a good measure for the brightness
differences of the ``naked'' stars (i.e. without additional accretion
luminosity from the disk). As is evident from Fig.\ \ref{deltaz},
there are fewer and fewer binaries per unit $\Delta$Z with increasing $\Delta$Z.

At first sight, this finding seems to be in contradiction to what Reipurth \& 
Zinnecker (1993) found for the CTTS alone; however, this is not so:
the only difference is that 
we have plotted brightness differences (in magnitudes) instead of 
flux ratios. Of course, in the end we are interested in the distribution
of mass ratios rather than brightness differences or flux ratios.

We also show in Fig.\ \ref{deltaz} a rough estimate on the corresponding mass 
ratios {\it (upper abscissa)} assuming a mass--luminosity
relation L$\propto$ M$^{1.6}$ for solar-type PMS stars on purely vertical
Hayashi tracks (T$_{\rm eff}$ = const.), 
 neglecting accretion luminosity and deuterium core and
shell burning (Brandner 1992, Zinnecker et al.\ 1992).

Preferentially large brightness differences between the individual components
of these wide binaries would imply large differences 
between primary and secondary mass. This puts an
 interesting constraint on theories of binary formation, as it is
still unknown which process plays the dominant role.
Although capture can be ruled out as a major source for binary star
formation (Clarke \& Pringle 1991), there remain at least three
viable formation processes: fragmentation of rotating disks (Shu et al.\ 1990),
fragmentation of a rotating filament (Zinnecker 1991, Bonnell et al.\ 1991) and collisional
fragmentation (Pringle 1989). 
While the fragmentation of rotating disks
should yield preferably small mass ratios  between the secondary
and the primary, the fragmentation of a rotating filament probably can
lead to rather equal mass ratios and thus to $q$ close to unity for
wide binaries.
The collisional
fragmentation is a more erratic process and so one expects more or less
uncorrelated masses. 

Mainly because of the uncertainties in our transformation from
brightness differences to mass ratios and because
IMFs based on observations of the luminosity function
(Miller \& Scalo 1979, Kroupa et al.\ 1990) are not corrected for
``contaminations'' by unresolved binaries (see Kroupa et al.\ 1991), 
at the moment we cannot favour one formation scenario over the others. 
Both collisional fragmentation and fragmentation of a rotating filament
would produce a mass-ratio distribution in agreement with our
observation. 
Only fragmentation of rotating disks can be ruled out with some
certainty as being the major formation mechanism for wide pairs
(but it still might be valid for close pairs, cf.\ Bonnell \& Bate 1994).

\section{Multiplicity}

\subsection{Binary frequency among G to M dwarfs}

\begin{figure}
\centerline{\psfig{figure=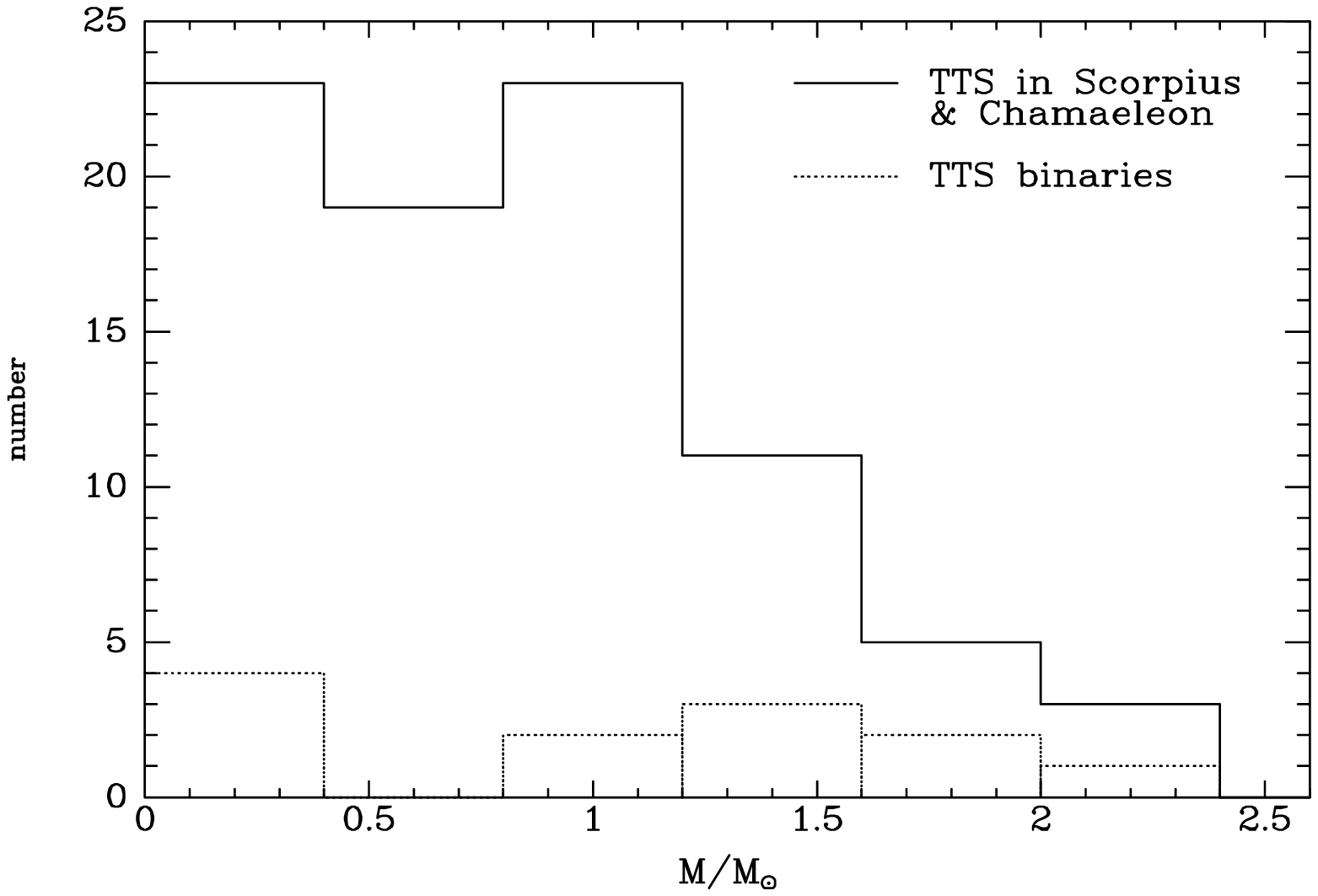,width=9.5cm}}
\caption{\label{mass}Histogram of the mass distribution of 84 X-ray selected
TTS in Scorpius and Chamaeleon in bins of 0.4 M$_\odot$ (solid line).
For comparison we also show the mass distribution of those TTS which
were resolved into binaries in the present survey (dotted line).
}
\end{figure}

Systematic surveys for companions to late-type main-sequence stars in the 
solar neighbourhood have been carried out by Duquennoy \& Mayor (1991)
and Fischer \& Marcy (1992). While the latter found 42\%$\pm$9\% of all M type
dwarfs to have one or more companions, the former found that 57\%$\pm$9\%
of all G type dwarfs possess at least one companion. Furthermore, 
Fischer \& Marcy could show that the binary period (and semimajor axis) 
distribution of the M dwarfs is not greatly different form that of the
G dwarfs within the error bars. 

Therefore, when comparing our late-type pre-main sequence binaries
with main-sequence binaries we mainly have to adjust for
the binary frequency varying with spectral type (mass).
Alcal\'a (1994) and Kunkel (1995) placed many of the TTS in Chamaeleon
and Scorpius on the HR diagram and derived masses from theoretical pre-main
sequence evolutionary tracks from D'Antona \& Mazzitelli (1994). 
Figure \ref{mass} shows the mass distribution
of those 84 TTS 
%(among the 169 X-ray selected TTS surveyed with SUSI/NTT) 
for 
which such mass determinations are available
(Alcal\'a 1994, Kunkel 1995). The assumption here is that all TTS are single 
stars. 
%{\it The fact that for only half of the stars in our sample mass determinations
%are available does not impose an additional selection bias. It is simply
%because unfavourable weather conditions prevented luminosity determinations.}

Almost half of the TTS are less massive than 0.8 M$_\odot$. Hence, it seems 
appropriate to assume a binary frequency intermediate between M and G type 
stars: 50\%$\pm$9\%. 
From period distribution \'a la Duquennoy \& Mayor
we then get a binary frequency of 13.5\%$\pm$3.0\% in a range of separations
between 120AU and 1800AU (see error discussion in section 7.3.).

\subsection{Binary frequency among X-ray and H$\alpha$ selected TTS}

\begin{table*}
\caption{\label{mult}Binary frequency (BF) of TTS in OB and T associations with
binary separations between 120AU and 1800AU. The corresponding main
sequence binary frequency is 13.5\%$\pm$3.0\%.}
\begin{center}
\begin{minipage}{175mm}
\begin{tabular}{l|c|c|c|c|c|c|c}
& \multicolumn{2}{c|}{WTTS} & \multicolumn{2}{c|}{CTTS} & \multicolumn{2}{c}{WTTS+CTTS}&\\
region& number &binaries (BF in \%) & number &binaries (BF in \%)& number &binaries (BF in \%)& reference\\ \hline
Chamaeleon&95 &6+1.5\footnote{Corrected for X-ray bias, incompleteness, and chance projections (see text) and normalized to the total number of TTS observed} (7.9$\pm$2.9)& 100 & 18 (18.0$\pm$4.2) & 195 & 25.5 (13.1$\pm$2.6)&BAKMZ, B, RZ \\ 
Upper Scorpius&74  &13-1.5$^a$ (15.5$\pm$4.6)  &--- &---      &74 & 11.5 (15.5$\pm$4.6)&BAKMZ \\ 
Lupus  &---&---&60  &9 (15.0$\pm$5.0)&60  &9 (15.0$\pm$5.0)&B, RZ \\
$\rho$ Ophiuchi&---&---&92  &13 (14.1$\pm$3.9)&92  &13 (14.1$\pm$3.9)&B, RZ \\ 
Taurus-Auriga  &44  &11 (25.0$\pm$7.5) &60 & 16 (26.7$\pm$6.7) &104 & 27 (26.0$\pm$5.0)&LZWCRJHL \\ \hline
total &213& 30 (14.1$\pm$2.6)& 312 & 56 (17.9$\pm$2.4) & 525 & 86 (16.4$\pm$1.8)\\ \hline
\end{tabular}  
References: B stands for Brandner (1992, 1995), BAKMZ for this work, 
LZWCRJH for Leinert et al.\ (1993), and RZ for Reipurth \& Zinnecker (1993).
\end{minipage}
\end{center}
\end{table*}

\subsubsection{Chamaeleon}

In total we have surveyed 95 X-ray selected TTS and 26 H$\alpha$ selected
TTS in Chamaeleon and 74 X-ray selected TTS in Scorpius
and Lupus. In the former region 14 binaries could be identified,
eleven of them having separations larger than 0\farcs8 (cf.\ Table \ref{cross}).
We get 
a binary frequency of 5.4\%$\pm$2.6\% in a range of separations
between 0\farcs8 and 12\farcs0 among X-ray selected T Tauri stars
in Chamaeleon. From a compilation of the lists of Brandner (1992, 1995),
Reipurth \& Zinnecker (1993), and this paper we get in total 100 CTTS (four
of them in common with our X-ray selected TTS) surveyed for multiplicity.
18 of these stars have a companion that could be confirmed as a T Tauri star 
in a range of separations between 0\farcs8 and 12$''$ (Brandner 1995). 
Thus the overall BF is (11-1.7+14)/(95-15.8+100) or 13.0\%$\pm$2.7\%, which is
in good agreement with the values for 
main sequence stars (13.5\%$\pm$3.0\%). Because of the less 
favourable seeing
conditions during the first third of our survey in Chamaeleon, we
could have missed 2 to 3 binaries with separations between
0\farcs8 and 1\farcs2. Also in the interval between 3$''$ to 12$''$
we have a couple of binary candidates which still need confirmation.
Hence,
one or two of the wider binaries might still be missing.
On the other hand, among the 95 studied
X-ray selected T Tauri stars we would expect one chance projection
and 2 additional binaries
because of the X-ray bias for the binaries with separations $\le$ 3$''$.
Two additional binaries
would give us a overall BF of 14.1\%$\pm$2.8\%, a value still in good
agreement with the main sequence value.

\subsubsection{Scorpius and Lupus}

In Scorpius and Lupus we found 13 binaries with separations between
0\farcs8 and 3$''$. As pointed out above, for the identification
of wider binaries we cannot rely on statistical arguments but need
additional information. We again have a number of
wider binary candidates which need confirmation by spectroscopy
and proper motion studies. Thus, for the time being, we can only give
some estimate on the number of binaries with separations between
3$''$ and 12$''$.

Similar to the Chamaeleon observations, we would expect 1 or 2 binaries more 
in the interval 3$''$\dots12$''$ -- given that we have the same distribution of 
separations among pre-main sequence binaries
as among main sequence binaries. We also expect 
one mere 
chance projection among the binaries with separations $\le$ 3$''$.
Including the correction for the X-ray bias,
we get a BF of (13-5)/(74-22.2) or 15.4\%$\pm$5.5\%, which is 
-- within the statistical uncertainties -- in agreement with the main 
sequence value.

\subsubsection{Summary}

Table \ref{mult} summarizes the binary frequency among WTTS and CTTS in Chameleon,
Upper Scorpius, Lupus, $\rho$ Ophiuchi, and Taurus  with binary separations in 
the range between 120 and 1800 AU. Looking at the total number of
TTS we find no significant difference in
the BF of WTTS and CTTS. 
There is a difference in the BF of the WTTS and CTTS in Cha.
However, the sample is small and the difference may not be statistically 
significant.
In a recent model for TTS with accretion disks,
Clarke et al.\ (1995) suggested  that a magnetic gate regulates
the accretion flow from the disk onto the star. Thus some stars
might interchange between being classified as  CTTS or WTTS, simply because
the ``magnetic gate'' is either open or closed. Such a mechanism would
help to smear out any existing difference between binary properties of 
CTTS and WTTS, if such a difference does indeed exist.

Without the Taurus-Auriga region, we get a BF of 14.0\% $\pm$ 1.8\% among TTS 
(compared to the main sequence BF of 13.5\% $\pm$ 3.0\%). 
If we include the Taurus-Auriga region, the 
BF of 16.4\% $\pm$ 1.8\% among TTS  is still not greatly different to
the main sequence BF.
It seems that only the Taurus-Auriga T association shows an enhanced
BF with respect to the main sequence and with respect to {\it other}
T and OB associations or clusters (Trapezium cluster). Why this is the case,
we do not know.
80 new TTS in Taurus identified by Wichmann (1994, see
also Wichmann et al.\ 1995) have not yet been surveyed for
multiplicity. Therefore, it remains to be
seen if the overall BF in Taurus-Auriga stays that high.

Similar to Reipurth \& Zinnecker (1993) in their study of CTTS, we find a low 
triple frequency among the X-ray selected TTS: only one out of the
150 TTS studied is a (candidate) visual triple. Given the limited range of separations
in our study, this is further evidence that hierarchical orbits are 
established early in PMS multiple systems.

We come to the conclusion that binarity itself is established very early in stellar evolution, probably even before the stars are crossing the stellar 
``birthline'' (Stahler 1983). Furthermore, there is no need
for any break-up of
binaries due to close encounters or any evolution of orbits, for
{\it wide} binaries with
separations larger than 120AU.

\subsection{Error discussion}

All the results on the BF of pre-main-sequence stars can only be referred to 
main-sequence values under some assumptions. This naturally leads to some
uncertainties propagating into our analysis of the BF.

In all five star forming regions surveyed by Alcal\'a (Chamaeleon \& Orion),
Wichmann (Lupus \& Taurus-Auriga), and Kunkel (Upper Scorpius \& the northern
part of Upper Centaurus--Lupus) WTTS could be found spread out over the
whole area under study.  No sharp boundaries of the star forming regions
could be found. This raises the interesting question whether the WTTS
are at all physically related to the
dark clouds and their CTTS. At least we must assume that the spread
in distance for the WTTS is comparable to their spatial extent perpendicular
to the line of sight. In Chamaeleon with the dark clouds Cha I and
II at a distance of 140pc and 170pc, respectively, and TTS covering a field
of $\approx$ 40pc $\times$ 30pc, this would mean that
each WTTS could be located anywhere at a distance between, say, 120pc and
190pc. The same holds true for the WTTS in Upper Scorpius and Lupus, and very likely also
for those in Taurus-Auriga and Orion. Thus derived luminosities for these
stars and hence the position in an HRD are uncertain by at least the
uncertainty in their distance estimate. The BF is not so much
affected by the uncertainty in the distance.

All transformation from orbital periods to separations are based on the
distribution of orbital elements for G type
stars in the solar neighbourhood (Duquennoy \& Mayor 1991).

\begin{itemize}

\item[i]{\bf uncertainty in distance estimates:}
The distance to the binaries is uncertain. As discussed in the text we get
a distance estimate of 150pc$\pm$30pc. This introduces an
additional scatter in our transformation of the main sequence BF
of $\pm$0.5\%.

\item[ii]{\bf transformation: observed separations -- semi--major axes}

Orbits show a random orientation in space. Because of the long
orbital periods of the wide binaries we have no a priori information on the
orbital parameters. From van Albada (1968) we get: $$<\ln \frac{d}{a} > = \ln
(1+\sqrt{1-e^2}) - \sqrt{1-e^2}\mbox{,}$$ where $d$ is the observed separation,
$a$ the semi--major axis, $e$ the eccentricity, and $<>$ indicates the
 average over all possible orientations.
For an eccentricity
of e = 0.5, $<\ln \frac{d}{a} > = -0.056$. Thus on average the error
by setting the observed separation equal to the semi--major axes is about
5\%.

\item[iii]{\bf transformation: orbital periods -- semi--major axes}

The transformation from orbital periods to semi-major axes and
vice versa seems to be the main uncertainty. Up to now we do not
have good mass estimates for primaries and secondaries, which
would be necessary for an accurate transformation for each individual
system. For the transformation from orbital periods to semi-major
axes we assumed a system mass of 1.5M$_\odot$ and circular orbits.
The system mass of 1.5M$_\odot$ would, e.g., correspond to a K0V (0.8M$_\odot$)
-- K5V (0.7M$_\odot$) pair, which we assume to be the ``typical''
pair (cf. Table \ref{cross} for spectral types and Fig. \ref{deltaz} for
the distribution of mass ratios)

\end{itemize}

Summing up, the overall uncertainties in the binary frequency are small
enough not to affect our conclusions.

\section{Summary}

We have surveyed 26 H$\alpha$ selected T Tauri stars in Chamaeleon and
169 X-ray selected T Tauri stars in Chamaeleon and
Upper Scorpius for companions. Our study is complete for separations 
down to 0\farcs8. In total we identified 31 binaries, 
twelve of them having separations less than 1\farcs0.

In both major regions we find the binary frequency 
(WTTS \& CTTS in Chamaeleon, in Scorpius 
WTTS only) in good agreement with the binary frequency among
main sequence stars in the same range of separations. Previous studies
of the Lupus and $\rho$ Ophiuchi T associations led to the same result.
From a
total of 525 T Tauri stars in the various associations (including
Taurus-Auriga), 86
possess companions
in a range of separations between 120AU and 1800AU. Hence, the overall 
binary frequency is
16.4\%$\pm$1.8\% among TTS, and does not greatly differ from
the binary frequency among
low-mass main sequence stars (13.5\%$\pm$3.0\%).
Only the Taurus-Auriga T association shows an enhanced binary frequency
with respect to other T and OB associations and to main sequence stars
in the solar neighbourhood. If we omit the Taurus-Auriga region in our analysis
we end up wih a binary frequency of 14.0\%$\pm$1.8\% which is in very good
agreement with the main sequence value.

We conclude that {\it wide} PMS binaries do not differ from main sequence
binaries with respect to orbital parameters and binary frequency. Thus
binary properties are established very early on in stellar evolution
and orbital parameters remain unchanged afterwards.

There is no preponderance for large brightness differences between primary and
secondary of individual systems. Hence, fragmentation of rotating disks
plays only a subordinate role in the formation of wide binaries.
Fragmentation of rotating filaments and collisional fragmentation are viable 
formation mechanisms for wide binaries.

\acknowledgements
We are grateful to Bo Reipurth for obtaining part of the NTT/SUSI data
and for comments and suggestions. We thank Jerome Bouvier for his
constructive referee's comments.
WB was supported by a student fellowship of the European Southern
Observatory and by the Deutsche Forschungsgemeinschaft (DFG) under grant
Yo 5/16-1. MK \& HZ acknowledge support by the DARA under grant 05 OR 9103 0.
The ROSAT project was supported by the
Bundesministerium f\"ur Forschung und Technologie (BMFT/DARA) and the
Max-Planck-Society.
This research has made use of the Simbad database,
operated at CDS, Strasbourg, France, and NASA's Astrophysics Data System (ADS),
version 4.0.


\begin{thebibliography}{}

\bibitem[]{} van Albada T.S. 1968, Bull. Astr. Inst. Netherlands 20, 47

\bibitem[]{} Alcal\'a J.M. 1994, PhD thesis, Ruprecht-Karls-Universit\"at Heidelberg

\bibitem[]{} Alcal\'a J.M., Krautter J., Schmitt J.H.M.M., Covino E., Wichmann 
R., Mundt R. 1995 A\&A, submitted

\bibitem[]{} Bastian U., R\"oser S. 1993, PPM star catalogue, Spektrum, Akad.\
Verl.\ Heidelberg, Berlin

\bibitem[]{} Bertout C., Basri G., Bouvier J. 1988, ApJ 330, 350

\bibitem[]{} Blaauw A., in {\it The Physics of Star Formation and Early Stellar Evolution}, eds. C.J. Lada \& N.D. Kylafis, ASI C 342, 125

\bibitem[]{} Bonnell I.A., Bate M.R. 1994, MNRAS 271, 999

\bibitem[]{} Bonnell I., Martel H., Bastien P., Arcoragi J.-P., Benz W. 1991,
ApJ 377, 553

\bibitem[]{} Bouvier J.\ 1990 AJ 99, 946

\bibitem[]{} Brandner W. 1992, Diploma thesis, Julius-Maximilians-Universit\"at W\"urzburg

\bibitem[]{} Brandner W. 1993, 5th ESO/ST-ECF Data Analysis Workshop, eds. Grosb{\o}l
\& de Ruijscher, p. 137

\bibitem[]{} Brandner W. 1995, PhD thesis, Julius-Maximilians-Universit\"at W\"urzburg, in preparation

\bibitem[]{} Clarke C.J., Pringle J.E. 1991, MNRAS 249, 588

\bibitem[]{} Clarke C.J., Armitage P.J., Smith K.W., Pringle J.E. 1995, 
MNRAS 273, 639

\bibitem[]{} Covino E., et al.\ 1995, A\&A in preparation

\bibitem[]{} Duquennoy A., Mayor M. 1991 A\&A 248, 485

\bibitem[]{} Durisen R.H., Sterzik M.F. 1994, A\&A 286, 84

\bibitem[]{} Feigelson E.D., Casanova S., Montmerle T., Guibert J. 1993,
ApJ 416, 623

\bibitem[]{} Feigelson E.D., Kriss G.A. 1981, ApJ 248, L35

\bibitem[]{} Fischer D.A., Marcy G.W. 1992 ApJ 396, 178

\bibitem[]{} de Geus E.J., de Zeeuw P.T., Lub J. 1989, A\&A 216, 44

\bibitem[]{} Ghez A.M., Neugebauer G., Matthews K. 1993, AJ 106, 2005

\bibitem[]{} Harrington R.S. 1975, AJ 80, 1081

\bibitem[]{} Hartigan P., Kenyon S.J., Hartmann L.W., Strom S.E.,
Edwards S., Welty A.D., Stauffer J. 1992, ApJ 382, 617

\bibitem[]{} Hartigan P. 1993, AJ 105, 1511

\bibitem[]{} Huenemoerder D.P., Lawson W.A., Feigelson E.D. 1994, MNRAS 271, 967

\bibitem[]{} Jefferys W.H., McArthur B., McCartney J.E. 1991, BAAS 23(2), 997

\bibitem[]{} Krautter J.\ et al.\ 1996, A\&A in preparation

\bibitem[]{} Kroupa P., Tout C.A., Gilmore G. 1990, MNRAS 244, 76

\bibitem[]{} Kroupa P., Tout C.A., Gilmore G. 1991, MNRAS 251, 293

\bibitem[]{} Kunkel M. 1995, PhD thesis, Julius-Maximilians-Universit\"at W\"urzburg

\bibitem[]{} Leinert Ch., Zinnecker H., Weitzel N., Christou J., Ridgeway S.T.,
Jameson, R., Haas, M., Lenzen, R 1993, A\&A, 278, 129 

\bibitem[]{} Mathieu R.D., Walter F.M., Myers P.C. 1989, AJ 98, 987

\bibitem[]{} Mathieu R.D. 1994, ARA\&A 32, 465

\bibitem[]{} Miller G.E., Scalo J.M. 1979, ApJS 41, 513

\bibitem[]{} Montmerle T., Koch-Miramond L., Falgarone E., Grindlay J.E. 1983,
ApJ 269, 182 

\bibitem[]{} Mundt R., Walter F.M., Feigelson E.D., Finkenzeller U., Herbig
G.H., Odell A.P. 1983, ApJ 269, 229

\bibitem[]{} Pringle J.E. 1989, MNRAS 239, 361

\bibitem[]{} Reipurth B., Graham J.A.\ 1991, in {\it Low Mass Star Formation
in Southern Molecular Clouds}, ed.\ Reipurth, ESO Scientific Report 11,
p.\ 149

\bibitem[]{} Reipurth B., Zinnecker H. 1993, A\&A, 278, 81

\bibitem[]{} Prosser C.F., Stauffer J.R., Hartmann L., Soderblom D.R., Jones B.F., 
Werner M.W., McCaughrean M.J. 1994 ApJ 421, 517

\bibitem[]{} Richichi A., Leinert Ch., Jameson R., Zinnecker H. 1994, A\&A 287, 145

\bibitem[]{} Scheffler H., Els\"asser H. 1965, Landolt-B\"ornstein,
ed. K.H. Hellwege, p. 604

\bibitem[]{} Schwartz R.D. 1977, ApJS 35, 161

\bibitem[]{} Schwartz R.D. 1989, in {\it Low Mass Star Formation
in Southern Molecular Clouds}, ed.\ Reipurth, ESO Scientific Report 11,
p.\ 93

\bibitem[]{} Shu F.H., Tremaine S., Adams F.C., Ruden S.P. 1990, ApJ 370, L31

\bibitem[]{} Simon M. 1992, in {\it Complementary Approaches to Double and
Multiple Star Research}, eds. H.A. McAlister \& W.I. Hartkopf, IAU Coll. 135, 41

\bibitem[]{} Simon M., Chen W.P., Howell R.R., Slovik D. 1992, ApJ 384, 212

\bibitem[]{} Stahler S.W. 1983, ApJ 274, 822

\bibitem[]{} Sterzik M.F., Alcal\'a J.M., Neuh\"auser R., Schmitt J.H.M.M.
1995, A\&A 297, 418

\bibitem[]{} Turon C., Cr\'ez\'e M., Egret D., et al.\ 1992, {\it The
Hipparcos Input Catalogue}, ESA, published in electronic form

\bibitem[]{} Walter F.M., Kuhi L.V. 1981, ApJ 250, 254

\bibitem[]{} Walter F.M., Vrba F.J., Mathieu R.D., Brown A., Myers P.C.
1994, AJ 107, 692

\bibitem[]{} Whittet D.C.B., Assendorp R., Prusti T., Roth M., Wesselius P.R.
1991, A\&A 251, 524

\bibitem[]{} Wichmann R. 1994, PhD thesis, Ruprecht-Karls-Universit\"at
Heidelberg

\bibitem[]{} Wichmann R., Krautter J., Schmitt J.H.M.M., et al.\ 1995, A\&A 
in preparation

\bibitem[]{} Wilking B.A., Schwartz R.D., Blackwell J.H. 1987, AJ 94, 106

\bibitem[]{} Worley C.E., Douglass G.G. 1984, {\it The Washington Double star
catalog}, US Naval Obs., published in electronic form

\bibitem[]{} Zinnecker H. 1991, IAU Symp. 147, 526

\bibitem[]{} Zinnecker H., Brandner W., Reipurth B. 1992, in 
{\it Complementary Approaches to Double and Multiple Star Research},
eds. H.A. McAlister \& W.I. Hartkopf, IAU Coll. 135, p. 50

\bibitem[]{} Zinnecker et al.\ 1996, A\&A in preparation

\end{thebibliography}
\end{document}